\documentclass[seceq]{ptptex}

\usepackage{graphicx}

\newcommand{\la}{\langle}
\newcommand{\ra}{\rangle}

\newcommand{\e}{\mathrm{e}}
\newcommand{\beq}{\begin{eqnarray}}
\newcommand{\eeq}{\end{eqnarray}}
\newcommand{\sbeq}{\begin{subeqnarray}}
\newcommand{\seeq}{\end{subeqnarray}}

\newcommand{\bl}{\biggl}
\newcommand{\br}{\biggr}
\newcommand{\bfk}{\mbox{{\boldmath $k$}}}

\newcommand{\bfv}{\mbox{{\boldmath $v$}}}
\newcommand{\bfr}{\mbox{{\boldmath $r$}}}

\newcommand{\bfq}{\mbox{{\boldmath $q$}}}

\title{
Dynamical Density Fluctuations around QCD Critical Point Based on
Dissipative Relativistic Fluid Dynamics %
}

\subtitle{possible fate of Mach cone at the critical point
\footnote{The preliminarily version of this work has been reported in Ref.\citen{report}}}    
\author{
Yuki \textsc{ Minami} and Teiji \textsc{ Kunihiro}%
}

\inst{
Department of Physics,
Kyoto University, Kyoto 606-8502, Japan
}

\abst{
The purpose of this paper is twofold.
Firstly, we study the dynamical density fluctuations around the
critical point(CP) of  Quantum Chromodynamics(QCD)
using dissipative relativistic fluid dynamics in which
the coupling of the density fluctuations to those of other conserved quantities is
taken into account.
We show that the sound mode which is directly coupled to the 
mechanical density fluctuation is attenuated 
 and in turn the thermal mode becomes the genuine soft mode 
at the QCD CP.
We give a speculation on the possible fate of a Mach cone
in the vicinity of the  QCD CP as a signal of the existence of the CP 
on the basis of the above findings. 
Secondly, we clarify that 
the so called first-order 
relativistic fluid dynamic equations have generically no problem
to describe fluid dynamic phenomena with long wave lengths contrary to
a naive suspect 
whereas even Israel-Stewart equation, a popular second-order equation,
may not describe the hydrodynamic mode in general depending
on the value of the relaxation time.    
}

\PTPindex{231,519}  

\begin{document}

\maketitle

\section{Introduction}

The analysis\cite{Hirano:2005xf}
 of the experimental data\cite{Arsene:2004fa} obtained in 
Relativistic Heavy Ion Collider (RHIC) at Brookhaven National Laboratory(BNL) has suggested
 that the matter created by the relativistic heavy ion collisions is well 
described as almost ideal relativistic fluids with
 the ratio of  shear viscosity $\eta$ 
to  entropy density $s$ being tiny.
This discovery has prompted
great interest in the origin and the true value of the viscosities including
the bulk viscosity $\zeta$ and so on, 
and people are now enthusiastic 
in elucidating dissipative effects\cite{dissipative} and thus 
in constructing a transport theory for strongly coupled 
systems\cite{tko,Baier:2007ix,Huovinen:2006jp}. 

A unique feature of the phase diagram of Quantum Chromodynamics(QCD)
 is the existence of a critical point(CP), which is yet mainly
 based on some effective models\cite{qcdcp} and a few lattice studies\cite{Ejiri:2007ga};
see Refs.\citen{Zhang:2008wx},\citen{Hatsuda:2006ps} and \citen{Fukushima:2008is} 
for possible variants and alternatives.
At the QCD CP, the first order phase transition terminates and turns to 
a second order phase transition.  Around a critical point of a second order transition,
 we can expect  large fluctuations of various quantities,
 and more importantly  there should exist a soft mode associated 
to the CP of second order\cite{Stephanov:1999zu}.
What is, however,  the soft mode of the QCD CP?
It has been established now that 
the QCD CP belongs to the same universality class as the liquid-gas CP
 and  shown that the density fluctuating mode 
and generically hydrodynamic
modes coupled to conserved quantities
 in the space-like region 
are a softening mode at the CP\cite{fujii,son}:
The would-be soft mode, the $\sigma$ mode, is coupled to the 
density fluctuation\cite{Kunihiro:1991qu} and becomes a 
slaving mode of the density variable\cite{fujii,son,Stephanov:2008qz};
 see Ref.~\citen{Ohnishi:2005br} for 
another argument on the fate of the $\sigma$ mode around the CP.
Furthermore it is also suggested \cite{karsch,Moore}
that the bulk viscosity   should show a divergent behavior
around  the QCD CP.\footnote{It seems that there exist some
refutable arguments in Ref.\citen{karsch}, as shown in
\citen{Moore,Romatschke:2009ng}.}

In this paper, we shall show that the
ultimate soft mode at the QCD CP may not
be a sound mode that is directly related to the 
dynamical density fluctuation, and
that the possible divergent behavior of the viscosities might not
be observed through the density fluctuations;
the sound modes are attenuated around the CP and would eventually
 die out at the CP; in turn,
the diffusive thermal mode that is coupled to the sound mode would
become the soft mode at the QCD CP.
We should mention that Fujii and Ohtani\cite{fujii}
emphasized that the thermal mode as well as the density oscillation
play the role of the soft modes in the QCD CP on the basis of the time-dependent
Ginzburg-Landau formalism.
However, the relative importance of these hydrodynamic modes around the QCD CP
was not elucidated.

Our investigation is  based on an explicit use
of the relativistic fluid dynamic
equations for a viscous fluid.
We shall show that even the so called first-order 
relativistic fluid dynamic equations have generically no problem
to describe fluid dynamical phenomena with long wave lengths contrary to
a naive suspect.

In the case of nonrelativistic fluids,
the nature of the critical dynamics around the critical point
is rather thoroughly elucidated\cite{stanley,kawasaki,onuki}.
The static density fluctuation induces a strong light scattering
in the vicinity of the critical point, which is 
called critical opalescence.
The dynamical density fluctuations are investigated 
on the basis of the (non-relativistic) Navier-Stokes
equation\cite{lp,stanley,onuki,reichl}. Note that the fluid dynamics is a dynamical
thermodynamics, which tells us that 
 the density fluctuation is in general coupled to the
entropy fluctuations creating diffusive thermal mode.

We shall explore how the singularities of 
the thermodynamic
 values as well as the transport coefficients affect the dynamical density
 fluctuations around the QCD CP using relativistic fluid dynamics for
a viscous fluid. We examine the spectral function 
of the density fluctuation in the equilibrium.
Our analysis is actually an extension of that  made for non-relativistic 
case\cite{lp,reichl} to the relativistic case. 
We are not aware of such an analysis so far
except for a simple estimate\cite{maartens} using Euler equation without dissipation
in a astronomical context.
The use of the relativistic fluid dynamics for  describing the
dynamics around QCD CP should 
be interesting in view of the success of the fluid dynamics for RHIC
phenomenology and the nature of the QCD CP mentioned above\cite{Paech:2003fe}. 
Admittedly, the analysis based on  fluid dynamics is 
only valid in the fluid dynamical regime: $k\xi << 1$,
where $k$ and $\xi$ denote the typical wave number of the fluid
and the correlation length;
that is, our discussions on the critical behavior are
based on  extrapolation from the fluid dynamical to the 
critical region: $k\xi >> 1$.\cite{HALPERIN:1969zza}
This extrapolation is known to make good sense 
for the non-relativistic case\cite{stanley}.

A set of the dynamical variables is thus the density $n$, temperature $T$
and the fluid velocity $u^{\mu}$ ($u_{\mu}u^{\mu}=1$; $\mu=0,\,1,,,\,3$),
while the sigma mode is not included as a dynamical variable because it is a slaving 
mode of the density\cite{fujii,son}.
We shall see that the coupling of the density fluctuation with that of the entropy
gives rise to an important effect and is found essential for the description
of the dynamics around the CP:
It will turn out that  the relativistic effects on the spectral function 
of the density fluctuation appears only in the width of the Lorentzian peaks
due to the sound and thermal modes
through the modification of the transport coefficients due to relativistic effects.
Moreover, our detailed analysis of the critical behavior of the transport coefficients as
well as the thermodynamic quantities will show that the possible singular  
behavior of the transport coefficients with relativistic effects
turns out to be masked by the more singular behavior of the specific heats;
hence  we find that the sound mode will be attenuated , and 
only the Rayleigh peak due to the thermal fluctuation stays out around the QCD CP,
which is found, to our surprise, precisely the same in effect as
in  the nonrelativistic case.\cite{stanley}

The forms of the 
 dissipative relativistic fluid dynamic equations are far from 
being established, although a great development
has been seen in a couple of years\cite{tko,Huovinen:2006jp}.
Relativistic fluid dynamic equations are 
the balance equations for energy-momentum and particle number,   
\begin{eqnarray}
\partial_\mu T^{\mu \nu}=0,  \label{eq:T}\\
\partial_\mu N^\mu =0,    \label{eq:n} 
\end{eqnarray}
where $T^{\mu \nu}$ is the energy-momentum tensor
and  $N^\mu$  the particle current.
They are expressed as
\begin{eqnarray}
T^{\mu \nu}&=&(\epsilon+P)u^{\mu}u^{\nu}-Pg^{\mu\nu}+\tau^{\mu\nu},\\
N^\mu &=& n u^\mu+\nu^\mu,    
\end{eqnarray}
where $\epsilon$ is the energy density, $P$  the pressure, $u^\mu$ the flow velocity, 
and $n$ the particle density, the dissipative part of the energy-momentum
tensor and the particle current are denoted by $\tau^{\mu \nu}$ and $\nu^\mu$,
respectively.

The so called first order equations such as Landau\cite{landau} and Eckart\cite{eckart}
equations are parabolic and formally violates the causality,
and hence called acausal.
Moreover, the Eckart equation which is defined for the particle frame
where the particle current does not have a dissipative part shows a pathological
 property that the fluctuations around the thermal equilibrium is
unstable\cite{hiscock}, while the Landau equation defined for the energy frame does
not show such a pathological behavior. 
The causality problem is circumvented in the Israel-Stewart equation\cite{is},
which is a second-order equation with relaxation times incorporated.
One should, however, note
that the problem of the causality is only encountered
when one tries to describe phenomena with fast velocities which 
is beyond the scope of the fluid dynamics;
nobody would actually expect that the fluid dynamics should describe phenomena
with a velocity comparable with or greater than 
the mean velocity $\bar{v}$ of the constituent particles,
though $\bar{v}$ is still less than the light velocity.
The fluid dynamic equations describe the behaviors of the conserved quantities 
in even larger scale than the mean free path.
Thus the phenomena which the fluid dynamics should describe
are  slowly varying ones with the wave lengths
much larger than the mean free path.
The problem of causality always takes place in short wavelength region such as shock wave phenomena\cite{shock},
so formally acausal fluid dynamic equations suffice in describing
fluid dynamical phenomena.
In fact we will see that the results for fluid dynamical modes
with long wave lengths are qualitatively the same 
irrespective whether the second-order or first-order equations are
used.

As for the instability seen in the Eckart equation, a new first-order
equation in the particle frame constructed by Tsumura, Kunihiro and Ohnishi (TKO)
\cite{tko} has no such a pathological behavior\cite{tk}. 
We employ Landau\cite{landau}, Eckart\cite{eckart} and Israel-Stewart(I-S)\cite{is}
 equation as typical equations,
and TKO equation.

This paper is organized as follows.
In Sec.2, we calculate the spectral function of the dynamical
density fluctuation around thermal equilibrium state. 
We show that relativistic effects on the spectral function of the density fluctuation
appears  only in the width of sound modes,
 irrespective of the choice of the frame.
In Sec.3 we analyze the behavior of the spectral function around the QCD CP with 
the scaling laws,
and show that the sound mode is attenuated  while
the thermal mode which is coupled to the density
fluctuation stands out  in the critical region.
Section 4 is devoted to discussions in which  the fundamental reason is
given why the sound mode is attenuated in the critical region in terms
of the correlation length $\xi$ which diverges at the critical point.
We  furthermore suggest that the attenuation of the sound mode
can be used to identify the existence of the QCD CP by the
relativistic heavy-ion collisions; a possible suppression or
 disappearance of a Mach cone can be such an example.
The final section is for summary and concluding remarks.
We give a detailed derivation of thermodynamic relations in Appendix A,
which are used in Sec.2.
In Appendix B, we  show explicitly how the pathological behavior in Eckart equation
affects the procedure of the calculation of the spectral function 
of the density fluctuations for an instructive purpose.
In Appendix C and D, we present
the detailed  derivation of the spectral function in TKO equation in the particle frame 
and Israel-Stewart(I-S) equation, respectively.

\section{Analysis of dynamical density fluctuation}

In this section, we derive the spectral function of the 
dynamical density fluctuations
using some typical relativistic fluid dynamic equations for a viscous fluid,
and discuss  relativistic effects on and the frame dependence of
the spectral function.
A detailed derivation of the spectral function is given for the Landau 
equation in the first subsection. Leaving the similar detailed derivations
to Appendix C and D, we present the results for the Tsumura-Kunihiro-Ohnishi(TKO)
equation\cite{tko} and the Israel-Stewart equation\cite{is} in particle frame in the
subsequent subsections.
       
\subsection{In the case of Landau equation (energy frame)}

In  Landau equation, the dissipative terms are given by
\begin{eqnarray}
\nu^{\mu} &=& \kappa \bl( \frac{n T}{w} \br)^2 \partial_{\perp}^\mu \bl(\frac{\mu}{T} \br),  \\
\tau^{\mu \nu}&=&\eta [\partial^{\mu}_{\perp}u^{\nu}
               +\partial^{\nu}_{\perp}u^{\mu}
               -\frac{2}{3}\Delta^{\mu\nu}(\partial_{\perp}{\cdot}u)]
               +\zeta\Delta^{\mu\nu}(\partial_{\perp}{\cdot}u),                          
\end{eqnarray} 
where $\eta$ is the shear viscosity, $\zeta$  the bulk viscosity, 
$\kappa$  the thermal conductivity, $T$  the temperature, $\mu$  the chemical potential,
and $w=\epsilon+P$ the enthalpy density.
 $\Delta^{\mu \nu}=g^{\mu \nu}-u^\mu u^\nu$ is the projection operator 
to the space-like vector and 
$\partial_{\perp}^{\mu}=\Delta^{\mu \nu}\partial_{\nu}$ 
the space-like derivative (gradient operator). 

We calculate the spectral function of the dynamical density fluctuation 
around the thermal equilibrium state by the linear approximation for the deviation 
from the equilibrium. The following calculational procedure 
is an extension of the non-relativistic case described in the text book 
Ref.\citen{reichl}.

Let us write 
 $n(x)=n_0+\delta n(x)$, 
$\epsilon(x)=\epsilon_0 + \delta \epsilon (x)$, $P(x)=P_0+\delta P(x)$, 
$\mu(x)=\mu_0 + \delta \mu (x)$, and $u^\mu (x)=u^\mu_0 + \delta u^\mu (x)$,
where the respective quantities in the equilibrium state
are denoted with $_0$.
For simplicity, let the equilibrium state be  the rest frame of the fluid,
$u_0^\mu=(1,\bf{0})$; then owing to the relation
$u_0^{\mu}\delta u_{\mu}=0$, it is found  that 
$\delta u_{\mu}$ takes the form $\delta u^\mu (x)= (0, \delta\bfv (x))$
with $\delta\bfv (x)$  to be determined together with other quantities
like $\delta n$ etc. 
Then  Landau equation Eqs.(\ref{eq:T}) and  (\ref{eq:n}) are reduced to
\begin{eqnarray}
\frac{\partial \delta n}{\partial t}+n_{0}\nabla\cdot\delta\bfv 
   =\kappa\frac{n_0}{w_0}[\frac{T_0}{w_0} \nabla ^2(\delta P)
   -\nabla ^2 (\delta T)], 
\label{eq:ncon} \\
w_{0}\frac{\partial \delta\bfv }{\partial t}-\eta\nabla^{2}\delta\bfv 
   -(\zeta+\frac{1}{3}\eta)\nabla(\nabla\cdot\delta\bfv )+\nabla(\delta P)=0,
\label{eq:pcon} \\
n_0\frac{\partial \delta s}{\partial t}
   +\frac{\kappa}{w_0}\nabla^2(\delta P)
   -\frac{\kappa}{T_0}\nabla^2(\delta T)=0,
\label{eq:spro} 
\end{eqnarray}  
in the  linear approximation. Here 
we have used the following thermodynamic relations.
\begin{eqnarray}
d\epsilon &=&T_0 d(n s)+\mu_0 d n,  \\
d P       &=& n_0 s_0 d T+n_0 d {\mu},  \\
w_0       &=& T_0 n_0 s_0+n_0{\mu}_0, 
\end{eqnarray}
where $s_0$ is the entropy per the particle number in the thermal equilibrium.
Note that the right hand side of Eq. (\ref{eq:ncon}) and 
the second term in Eq. (\ref{eq:spro}) represent relativistic effects
which are absent for the  non-relativistic Navier-Stokes equation.
In addition, the coefficient of the first term in Eq. (\ref{eq:pcon}) $w_0$
 is replaced by the mass density $\rho_0$ for the Navier-Stokes equation. 

Now we have five equations for seven unknown quantities, $\delta n$, $\delta T$,
$\delta P$, $\delta s$, and $\delta \bfv $.
In order to solve these equations,  the thermodynamic quantities $\delta n$, $\delta T$, 
$\delta P$ and $\delta s$ are interrelated using  thermodynamics.
Choosing $\delta n$ and  $\delta T$ as independent variables,  we have
\begin{eqnarray}
\delta P(x)=\frac{w_{0} c_{s}^{2}}{n_{0} \gamma}\delta n(x)
            + \frac{w_{0}c_{s}^{2}\alpha_{P}}{\gamma}\delta T(x),       \\
\delta s(x)=-\frac{w_{0}c_{s}^{2}\alpha_{P}}{n_{0}^{2}\gamma}\delta n(x)
            + \frac{\tilde c_{n}}{T_{0}} \delta T(x),
\end{eqnarray}
where the following thermodynamic identities are used,   
\begin{eqnarray}
\bl(\frac{\partial P}{\partial n}\br)_{T}&=&\frac{w_{0} c_{s}^{2}}{n_{0} \gamma}, \hspace{1.0cm} 
\bl(\frac{\partial P}{\partial T}\br)_{n}=\frac{w_{0}c_{s}^{2}\alpha_{P}}{\gamma},  \nonumber \\ 
\bl(\frac{\partial s}{\partial n}\br)_{T}&=&
                                  -\frac{w_{0}c_{s}^{2}\alpha_{P}}{n_{0}^{2}\gamma}, 
\;\; \bl(\frac{\partial s}{\partial T}\br)_{n}=\frac{\tilde c_{n}}{T_{0}}
\end{eqnarray}
where $\tilde{c}_n=T_0(\partial s / \partial T)_n$ and $\tilde{c}_P=T_0(\partial s /
 \partial T)_P$
are the specific heats at constant density and pressure, respectively, 
$c_s=(\partial P/\partial \epsilon)_s^{1/2}$ the sound velocity, 
$\alpha_P=$\, $-(1/n_0)(\partial n / \partial T)_P$  the thermal expansivity 
at constant pressure, and $\gamma=\tilde{c}_P/\tilde{c}_n$  the ratio of the specific
 heats.   
Then Eqs. (\ref{eq:ncon})-(\ref{eq:spro}) take the form
\begin{eqnarray}
\{ \frac{\partial}{\partial t}-\kappa\frac{T_0 c_s^2}{w_0 \gamma}\nabla^2 \} \delta n
 +n_0\nabla\cdot\delta\bfv 
 +\kappa\frac{n_0}{w_0}(1-\frac{c_s^2 \alpha_P T_0}{\gamma})\nabla^2 \delta T=0, \\
w_{0}\frac{\partial \delta\bfv }{\partial t}-\eta\nabla^{2}\delta\bfv 
 -(\zeta+\frac{1}{3}\eta)\nabla(\nabla\cdot\delta\bfv )
 +\frac{w_0 c_s^2}{n_0\gamma}\nabla \delta n
 +\frac{w_0 c^2_s \alpha_P}{\gamma}\nabla \delta T=0,                        \\
\{ -\frac{w_0 c_s^2 \alpha_P }{n_0 \gamma }\frac{\partial }{\partial t}
  +\kappa \frac{c_s^2}{n_0\gamma } \nabla^2  \} \delta n
 +\{ \frac{n_0\tilde{c}_n}{T_0} \frac{\partial }{\partial t}
  +\kappa(\frac{c_s^2\alpha_P }{\gamma }-\frac{1}{T_0})\nabla^2 \} \delta T=0. 
\end{eqnarray}

To utilize the correlations in the initial time to obtain the
correlation function at later time $t$, 
we perform Fourier-Laplace transformation 
\begin{equation*}
\delta n(\bfk ,z)=\int_{-\infty}^{+\infty}d\bfr \int^{\infty}_{0}dt
\: \e ^{- z t-i\bfk \cdot\bfr } \delta n(\bfr ,t),\cdots etc.
\end{equation*}
Then we find
\begin{eqnarray}
( z + k^2 \kappa\frac{T_0 c_s^2}{w_0 \gamma} ) \delta n (\bfk ,z)
 +i n_0 \bfk  \cdot \delta\bfv (\bfk ,z)
 +k^2 \kappa\frac{n_0}{w_0}(\frac{c_s^2 \alpha_P T_0}{\gamma}-1) \delta T  \nonumber \\ 
 =\delta n(\bfk ,t=0),          
\label{eq:fln}             \\
z w_{0}\delta\bfv (\bfk ,z)+k^2 \eta \delta\bfv (\bfk ,z)
 +(\zeta+\frac{1}{3}\eta)\bfk (\bfk  \cdot \delta\bfv (\bfk ,z)) 
 +i \bfk  \frac{w_0 c_s^2}{n_0\gamma}\delta n(\bfk ,z)   \nonumber \\
 +i\bfk  \frac{w_0 c^2_s \alpha_P}{\gamma} \delta T(\bfk ,z)
 =w_0 \delta \bfv (\bfk ,t=0),        
\label{eq:flm}             \\
(-z\frac{w_0 c_s^2 \alpha_P }{n_0 \gamma }
 -k^2 \kappa \frac{c_s^2}{n_0\gamma }   )\delta n(\bfk ,z)
 +\{ z \frac{n_0\tilde{c}_n}{T_0} 
 -k^2 \kappa(\frac{c_s^2\alpha_P }{\gamma }-\frac{1}{T_0}) \} \delta T(\bfk ,z) 
\nonumber \\
 =-\frac{w_0 c_s^2 \alpha_P }{n_0 \gamma }\delta n(\bfk ,t=0)
  +\frac{n_0 \tilde{c}_n}{T_0}\delta T(\bfk ,t=0).
\label{eq:fls}
\end{eqnarray}

It is convenient to divide  the velocity into longitudinal and transverse components
\begin{equation}
\delta\bfv (\bfk ,z)=\delta\tilde{v}_{\parallel }(\bfk ,z) \hat{\bfk }
                           +\delta\bfv _{\perp }(\bfk ,z). 
\end{equation}
The transverse component of Eqs.(\ref{eq:fln})-(\ref{eq:fls}) reads 
\begin{equation}
z w_{0}\delta\bfv_{\perp }(\bfk ,z)+k^2 \eta \delta\bfv_{\perp }(\bfk ,z)
 =w_0 \delta\bfv_{\perp } (\bfk ,t=0).
 \label{eq:transverse}
\end{equation}
We shall not treat this equation, which admits a diffusive solution,
in the present work
 since this equation is decoupled to the density fluctuation.
We note that a complete treatment based on the 
mode-mode coupling theory \cite{stanley,kawasaki,onuki} of the critical dynamics
in the close vicinity of the CP involves the diffusive transverse mode
 as well as the thermal one which is also
 diffusive as we will see. 
 
The longitudinal component of Eqs.(\ref{eq:fln})-(\ref{eq:fls}) can be cast into 
 the following matrix form
\begin{equation}
A  
  \begin{pmatrix}
  \delta n(\bfk ,z) \\
  \delta v_{\parallel}(\bfk ,z) \\
  \delta T(\bfk ,z)
  \end{pmatrix}
  =  
  \begin{pmatrix}
  \delta n(\bfk ,0) \\
  \delta v_{\parallel}(\bfk ,0) \\
  -\frac{\alpha_{P}c_{s}^{2}}{n_{0}\gamma}\delta n(\bfk ,0)
  +\frac{n_{0}\tilde{c}_{n}}{T_{0}w_{0}}\delta T(\bfk ,0)
  \end{pmatrix},
\label{eq:matrix}
\end{equation}
where the matrix A is defined by
\begin{equation}
A= 
  \begin{pmatrix}
  z + k^2 \kappa \frac{T_0  c_{s}^{2}}{w_0\gamma}  & i k n_{0} & 
  -k^2 \kappa \frac{n_0}{w_0}(1-\frac{\alpha_P c_{s}^{2}T_0}{\gamma})  \\
 i k \frac{c_{s}^{2}}{n_{0}\gamma} 
  & z+\nu_{l}k^{2} 
  & i k\frac{\alpha_{P} c_{s}^{2}}{\gamma}\\
 \frac{n_0 \tilde{c}_n}{w_0 T_0}
  [-z\frac{w_{0}T_{0}\alpha_{P}c_{s}^{2}}{n_{0}^{2}
    \tilde{c}_{n}\gamma}-k^{2}\chi c_{s}^{2}\frac{T_0}{n_0}] 
  & 0
  & \frac{n_0 \tilde{c}_n}{w_0 T_0}
    [z+k^{2}\gamma \chi (1-\frac{\alpha_{P}c_{s}^{2}T_{0}}{\gamma})]
\end{pmatrix}.
\end{equation}
Here, we have introduced the longitudinal kinetic viscosity $\nu_l$, and
the thermal diffusivity $\chi$, 
\begin{equation}  
\nu_l=(\zeta+\frac{4}{3}\eta)/w_0, \hspace{1.0cm} \chi=\frac{\kappa}{n_0 \tilde{c}_P}.
\label{eq:long-kin}
\end{equation}
Multiplying the inverse $A^{-1}$ of the matrix A from the left in Eq.(\ref{eq:matrix}),
we obtain the Fourier-Laplace coefficient of the density fluctuation  
\begin{eqnarray}
\label{eq:tilden}
 \delta n(\bfk ,z)=\{ (A^{-1})_{1 1} 
 -\frac{\alpha_{P}c_{s}^{2}}{n_{0}\gamma}(A^{-1})_{1 3}\} \delta n(\bfk ,0)
+(A^{-1})_{1 2} \delta v_{\parallel}(\bfk ,0)    \nonumber                   \\
+\frac{n_{0}\tilde{c}_{n}}{T_{0}w_{0}} (A^{-1})_{1 3} \delta T(\bfk ,0).
\end{eqnarray}

Now, we are interested in the spectral function of the dynamical density fluctuation 
\beq
S_{n n}(\bfk ,\omega) \equiv \la \delta n(\bfk ,\omega) \delta n (\bfk ,t=0)\ra,
\eeq
where $ \delta n(\bfk, \omega) $  is the Fourier transformation of the density fluctuation defined by
\begin{equation} 
\delta n (\bfk , \omega) = \int^{\infty}_{-\infty}d\bfr \int^{\infty}_{-\infty}d t
e^{-i\omega t -i\bfk \cdot \bfr} \delta n(\bfr, t),
\end{equation} 
and $\la\,\,\ra$ denotes the thermal average in the equilibrium. 
We see that $S_{n\, n}(\bfk ,\omega)$ is obtained  by taking the thermal average of
Eq.(\ref{eq:tilden})  multiplied by  $\delta n (\bfk ,0)$.
However,  note that $\delta T$ and $\delta n$ are statistically independent in fluids system
which is established in Einstein fluctuation theory\cite{reichl}, and accordingly
\beq
\la \delta T(\bfk ,0)\delta n(\bfk ,0)\ra =0,
\eeq
Similarly, 
\beq
\la \delta v_{\parallel}(\bfk ,0)\delta n(\bfk ,0)\ra =0.
\eeq
Hence, we have
\beq
S_{n n}(\bfk ,\omega)=\{ (A^{-1})_{1 1} 
 -\frac{\alpha_{P}c_{s}^{2}}{n_{0}\gamma}(A^{-1})_{1 3}\}
\la  \delta n(\bfk ,0) \delta n(\bfk ,0)\ra.
\eeq

The needed matrix elements $(A^{-1})_{1 1}$ and $(A^{-1})_{1 3}$ 
may be calculated using the simple formula
$(A^{-1})_{1 1}=\frac{1}{\det A}(A_{2 2}A_{3 3}-A_{2 3}A_{3 2})$
and $(A^{-1})_{1 3}=\frac{1}{\det A}(A_{1 2}A_{2 3}-A_{1 3}A_{2 2})$.
Here, $\det A$ reads 
\begin{eqnarray}
\det A =\frac{n_0 \tilde{c}_{n}}{w_0 T_0}
         [z^3 
          &+&z^2 k^2 \{ \gamma \chi +\nu_l +\kappa\frac{T_0 c_s^2}{w_0 }
         -2\chi c_s^2 \alpha_P T_0 \}                  \nonumber  \\
          &+&z k^2 c_s^2
          +k^4 c_s^2 \chi
          +\cdots  ],
\end{eqnarray}
where we have used the thermodynamic identity (\ref{eq:tid}), and 
'$\cdots$' denotes the higher order terms in 
$k$ which we assume small because we are interested in the fluid dynamical modes.
$\det A$ can be nicely factorized in this approximation,
\begin{equation}
\det A \sim \frac{n_0 \tilde{c}_n}{w_0 T_0}
(z+\Gamma_{R}k^{2})(z+\Gamma_B k^{2}+i c_{s}k)(z+\Gamma_B k^{2}-i c_{s}k),
\end{equation}
where 
\begin{eqnarray}
\Gamma_{\rm R}&=&\chi, \\
\Gamma_{\rm B}&=&\frac{1}{2}[\chi(\gamma-1)+\nu_{l}
+c_s^2 T_0 (\kappa / w_0 -2\chi \alpha_P )].
\end{eqnarray}

Then we can write the Fourier-Laplace coefficient of the density fluctuation
to second order in $k$ 
\begin{equation*}
\frac{\delta n(\bfk ,z)}{\delta n(\bfk ,0)} \sim 
\frac{(z+\Gamma_{\rm B} k^{2}+i c_{s}k)(z+\Gamma_{\rm B} k^{2}-i c_{s}k)
+z k^{2}\frac{\kappa c_s^2}{w_0\gamma}
-k^2\frac{c_s^2}{\gamma}}
{(z+\Gamma_{\rm R}k^{2})(z+\Gamma_{\rm B} k^{2}+i c_{s}k)(z+\Gamma_{\rm B} k^{2}-i c_{s}k) }.
\end{equation*}

Performing the inverse Laplace transformation 
\begin{equation}
\delta n(\bfk ,t)/\delta n(\bfk ,0)=\frac{1}{2\pi i} \int ^{\delta+i\infty }_{\delta-i\infty}dz
\e ^{zt}\delta n(\bfk ,z)/\delta n(\bfk ,0),
\end{equation}
we obtain the dynamical density fluctuation at $t$
\begin{equation}
\delta n(\bfk ,t)/\delta n(\bfk ,0) \sim  
(1-\frac{1}{\gamma})
\e^{-\Gamma_{\rm R} k^{2}t} + \frac{1}{\gamma}
\cos(c_s k t) \e^{-\Gamma_{\rm B} k^{2} t}. 
\label{eq:tdensity}
\end{equation}
Here, we have retained only the terms in the amplitudes to zeroth order in $k$.
Since Eq.(\ref{eq:tdensity}) is the density fluctuation in a stationary process, 
we can replace the time $t$ by $\vert  t \vert $.
Therefore the Fourier transformation of Eq.(\ref{eq:tdensity}) is given by
\begin{eqnarray}
\delta n(\bfk ,\omega )/\delta n(\bfk ,0)=
(1&-&\frac{1}{\gamma})
\frac{2\Gamma_{\rm R} k^{2}}{\omega^{2}+\Gamma_{\rm R}^{2}k^{4}}
 \nonumber \\
&+&\frac{1}{\gamma}
[\frac{\Gamma_{\rm B} k^{2}}{(\omega -c_{s}k)^{2}+\Gamma_{\rm B}^{2}k^{4}}
+\frac{\Gamma_{\rm B} k^{2}}{(\omega +c_{s}k)^{2}+\Gamma_{\rm B}^{2}k^{4}}].
\end{eqnarray}
Thus we finally obtain the spectral function of the density fluctuation  
\begin{eqnarray}
S_{n n}(\bfk ,\omega ) &=& \la \delta n(\bfk ,\omega ) \delta n(\bfk ,t=0)\ra \nonumber \\
   &=& \la (\delta n(\bfk ,t=0))^2\ra [\;(1-\frac{1}{\gamma})
   \frac{2\Gamma_{\rm R} k^{2}}{\omega^{2}+\Gamma_{\rm R}^{2}k^{4}}
   \nonumber \\
   &+&\frac{1}{\gamma}
   \{\frac{\Gamma_{\rm B} k^{2}}{(\omega -c_{s}k)^{2}+\Gamma_{\rm B}^{2}k^{4}}
   +\frac{\Gamma_{\rm B} k^{2}}{(\omega +c_{s}k)^{2}+\Gamma_{\rm B}^{2}k^{4}}\} \;].
   \label{eq:landau}
\end{eqnarray}
We see that the spectral function have three peaks at frequencies $\omega=0$ and 
$\omega = \pm c_s k$:
The peak at $\omega=0$ corresponds to thermally induced density fluctuations.
This peak is called the Rayleigh peak,  while
the two side peaks at $\omega=\pm c_s k$ correspond to mechanically induced density 
fluctuation, i.e. sound waves.
These two peaks are called Brillioun peaks.

Now let us  compare this results with that in the  nonrelativistic case\cite{lp, reichl};
\begin{eqnarray}
S^{\rm NR}_{n n}(\bfk ,\omega ) 
   = \la (&\delta n&(\bfk ,t=0))^2\ra [\;(1 - \frac{1}{\gamma})
   \frac{2\chi k^{2}}{\omega^{2}+\chi^{2}k^{4}}
   \nonumber \\
  &+&\frac{1}{\gamma}
   \{\frac{\Gamma_B^{\rm NR} k^{2}}{(\omega -c_{s}k)^{2}+
{\Gamma_{\rm B}^{\rm NR}}^{2}k^{4}}
   +\frac{\Gamma_{\rm B}^{\rm NR} k^{2}}
{(\omega +c_{s}k)^{2}+{\Gamma_{\rm B}^{\rm NR}}^{2}k^{4}}\} \;],
   \label{eq:navier}
\end{eqnarray} 
where 
\begin{eqnarray}
\Gamma_{\rm B}^{\rm N R} &&=\frac{1}{2}[\chi(\gamma-1)+\nu_{l}^{\rm N R}],\\
\label{eq:gnr}
\nu_{l}^{\rm N R} &&=(\zeta + \frac{4}{3}\eta) / \rho_0 .
\end{eqnarray}
We see that  relativistic effects 
appear only in the width of the Brillouin  peaks:
\beq
\Gamma_{\rm B}=\Gamma_{\rm B}^{\rm MR}+\delta \Gamma_{\rm B}^{\rm La}
\equiv \Gamma_{\rm B}^{\rm La},
\eeq
where
\begin{eqnarray}
\Gamma_{\rm B}^{\rm MR} &&\equiv \frac{1}{2}[\chi(\gamma-1)+\nu_l ],
\end{eqnarray}
with the longitudinal kinetic viscosity $\nu_l$ defined in (\ref{eq:long-kin}),
and
\begin{equation}  
\delta \Gamma_{\rm B}^{\rm La} \equiv \frac{1}{2}c_s^2 T_0 (\kappa / w_0 -2\chi \alpha_P ).
\end{equation}
Firstly, the longitudinal kinetic viscosity is expressed in terms of the enthalpy
density $w_0$ in the relativistic case in place of the mass density $\rho_0$.
We call this modification the minimal relativistic (MR) effect.
The other is a genuine relativistic effect $\delta \Gamma_{B}^{\rm La}$
which is absent in the non-relativistic case.
This part comes from the right hand side of Eq. (\ref{eq:ncon}) and 
the second term in Eq. (\ref{eq:spro}) which represent relativistic effects and 
vanishes if we take the light speed $c \rightarrow \infty$. 
We see that the width of the Rayleigh peak 
is the same as the non-relativistic case\cite{lp, reichl} and the relativistic
effects on the width of the sound modes tend to cancel with each other,
so even the relativistic effect on the Brillouin peaks may be moderate.

In order to see the relativistic effects $\delta \Gamma_{B}^{\rm La}$ quantitatively,
we calculate thermodynamical values,\, $c_s,\alpha_P$ and $\gamma$ by
the equation of state(EoS) of massless classical ideal gas, $\epsilon=3 P=3 n T$, 
so we have $\tilde{c}_p=4$, $\tilde{c}_n=3$, $\alpha_P = 1 / T_0$, $c_s = \sqrt{1/3}$ 
and $s_{0}=4 -  \mu_{0} /T_{0} $. 
Then the widths in the  minimal relativistic case are given by
\begin{eqnarray}   
\Gamma_{\rm R}&=& \frac{\kappa}{4 n_0}, \\
\Gamma_{\rm B}^{\rm M R} &=& \frac{1}{24 n_0 T_0}(4\eta+3\zeta+\kappa T_0),
\label{eq:nrwidth}
\end{eqnarray}
and the relativistic corrections are found to be
\begin{eqnarray}
\delta \Gamma_{\rm B}^{\rm La} &=& \frac{1}{2}c_s^2 T_0 (\kappa / w_0 -2\chi \alpha_P )
                                =-\frac{\kappa}{24 n_0}.
\end{eqnarray}
We see that the relativistic effects in the Brillouin peaks is 
solely due to the thermal conductivity and has a negative value,
implying that the Brillouin peaks
get enhanced with a smaller width by the relativistic effect.
It addition, this relativistic term exactly 
cancels out with the minimal relativistic part 
due to the thermal conductivity in Eq.(\ref{eq:nrwidth}); 
thus we see that 
the thermal conductivity does not affect the Brillouin peaks in net
for the  massless classical ideal gas.  

\begin{figure}[!t]
 \begin{minipage}{1.0\hsize}
  \centering
   \includegraphics[width=70mm,angle=270]{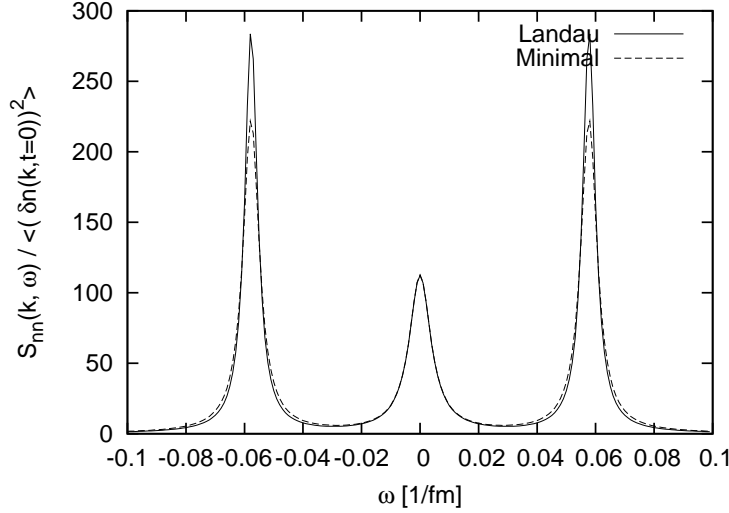}
  \caption{The spectral function in Landau  and the 
           minimal relativistic equation are shown by the solid line
           and the dashed line, respectively.
           The parameters are $k=0.1$[1/fm],\, $\mu_0=200$[MeV],\, 
           $T_0=200$[MeV],\, $\eta/(n_0 s_0)=\zeta/(n_0 s_0)=0.3$ and
           $\kappa T_0/(n_0 s_0)=0.6$. 
           Relativistic effects does not show up in the Rayleigh peak
 but enhance the Brillouin peaks.}
  \label{fig:landau}
 \end{minipage}
\end{figure}

Figure \ref{fig:landau} shows the spectral function Eq.(\ref{eq:landau}) and 
 the minimal relativistic case with 
the parameter set\, 
$k=0.1$[1/fm],\, $\mu_0=200$[MeV],\, 
$T_0=200$[MeV],\, $\eta/(n_{0}s_0)=\zeta/(n_0 s_0)=0.3$ and $\kappa T_0/(n_0 s_0)=0.6$.
Note that $n_0 s_0$ represents the entropy density in the equilibrium state 
because $s_0$ is the entropy per particle number.   

As is expected,  Fig.\ref{fig:landau} shows that
the Brillouin peaks owing to the sound mode
 is enhanced by the relativistic effects,
while the Rayleigh peak owing to the thermal mode 
is the same as in the non-relativistic case.

\subsection{In the case of particle frame}

Next let us take  fluid dynamic equations
in the particle frame. Does  any difference arise in the
density spectral function depending on the choice of the frame.
A typical equation in the particle frame is due to
Eckart\cite{eckart}.
It is, however, well known that the Eckart equation shows a
pathological behavior\cite{hiscock}; i.e,
when the  fluctuations around the thermal equilibrium is 
described by this equation, 
the fluctuation tends to diverge as $t$ goes infinity;
see Appendix B for a detailed discussion.

Then, does any first-order relativistic fluid dynamic equation for
a viscous fluid in the particle frame show such a pathological
behavior?
It is then noteworthy that a new fluid dynamic equation in the 
particle frame constructed
from the relativistic Boltzmann equation 
by Tsumura, Kunihiro and Ohnishi (TKO)\cite{tko} 
turns out to be a stable one in the sense that any fluctuations
around the thermal equilibrium state relax down to recover the 
equilibrium\cite{tk}.
So let us take  TKO equation in the particle frame and 
examine whether the spectral function of the density fluctuations
 shows any frame dependence.

In the case of TKO equation in the particle frame, the dissipative terms are given by
\begin{eqnarray}
\label{eq:tkoeq}
\tau^{\mu\nu}=&\eta &[\,\partial^{\mu}_{\perp}u^{\nu}+
   \partial^{\nu}_{\perp}u^{\mu}-\frac{2}{3}\Delta^{\mu\nu}(\partial_{\perp}{\cdot}u)\,]
   -\zeta^{'}(3 u^{\mu}u^{\nu}-\Delta^{\mu\nu})(\partial_{\perp}{\cdot}u) \nonumber \\
   &+&\kappa(u^{\mu}\partial^{\nu}_{\perp}T+u^{\nu}\partial _{\perp}^{\mu}), \\
\nu^{\mu}=&0&,
\end{eqnarray}
where $ \zeta^{'} =\zeta/(3\gamma-4)^2$ with $\gamma$ being the ratio of the 
specific heats as before. 

From the same procedure as taken for the Landau equation, 
we obtain the spectral function as follows
(the detailed derivation is given in Appendix C);
\begin{eqnarray}
\frac{S_{n n}(\bfk ,\omega )}{\la (\delta n(\bfk ,t=0))^2\ra }=
(1&-&\frac{1}{\gamma})\frac{2\chi k^{2}}{\omega^{2}+\chi^{2}k^{4}} 
\nonumber \\
&+&\frac{1}{\gamma}
[\frac{\Gamma_{\rm B} k^{2}}{(\omega -c_{s}k)^{2}+\Gamma_{\rm B}^{2}k^{4}}
+\frac{\Gamma_{\rm B} k^{2}}{(\omega +c_{s}k)^{2}+\Gamma_{\rm B}^{2}k^{4}}],
\label{eq:tko}
\end{eqnarray}
with
\begin{equation}
\Gamma_{\rm B}=\frac{1}{2}[\chi(\gamma-1)+\nu^{\rm TKO}_{l}
  -\frac{\alpha_{P}c_{s}^{2}}{n_{0}\tilde{c}_{P}}
(\kappa T_{0}+3\zeta^{'})] 
\equiv \Gamma_{\rm B}^{\rm TKO},
\end{equation}
where
\beq
\nu^{\rm TKO}_{l}\equiv (\zeta'+\frac{4}{3}\eta)/w_0.
\label{eq:nutko}
\eeq
Note that since the fluid dynamic  fluctuations around the equilibrium 
state is relaxing in TKO equation (see Appendix C), we have obtained the spectral
function without difficulty as was in the case of Landau equation.
Moreover, the deviation of the width of the Brillouin peaks reads
\beq
\delta \Gamma_{\rm B}^{\rm TKO} \equiv 
-\frac{\alpha_{P}c_{s}^{2}}{2 n_{0}\tilde{c}_{P}}(\kappa T_{0}+3\zeta^{'}),
\eeq
which is definitely negative. So the relativistic effect acts to 
enhance and sharpen the spectral function of the density fluctuation
in the TKO equation in the particle frame
in comparison with the minimal relativistic case than in 
the energy frame; see, however, the next subsection 
for the Israel-Stewart equation
in the particle frame.

If we estimate the relativistic effects using the EoS 
of massless classical ideal gas,
we have
\begin{equation}
\label{eq:deltatko}
\delta \Gamma_{\rm B}^{\rm TKO} 
=-\frac{1}{24 n_0 }(\kappa  + 3\zeta^{'}/T_0 ). 
\end{equation}
It is noted that the effective bulk viscosity $\zeta^{'}=\zeta/(3\gamma-4)^2$ is 
finite in the massless case
although $3\gamma-4 \rightarrow 0$.\cite{tko,tk}
The minimal relativistic part of the width is Eq.(\ref{eq:nrwidth}) with $\zeta$ 
replaced by $\zeta'$.
We see that the relativistic effect due to the thermal conductivity is 
the same as Landau equation
but another relativistic effect due to the bulk viscosity exists. 
In addition the relativistic effect cancels out
with  some part of  the minimal relativistic contribution
 due to the bulk viscosity and the thermal conductivity 
in Eq.(\ref{eq:nrwidth}) with 
$\zeta$ replaced by $\zeta'$.
That is, the bulk viscosity and the thermal conductivity does not have any contribution 
in net in TKO equation 
for the massless classical ideal gas.

\begin{figure}[!t]
 \begin{minipage}{1.0\hsize}
  \centering
   \includegraphics[width=70mm,angle=270]{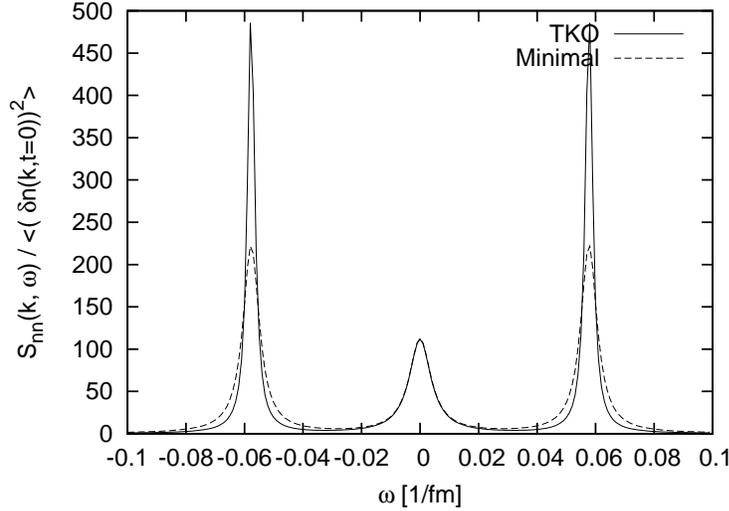}
  \caption{The spectral function in full relativistic TKO equation (solid line) and
its minimal relativistic version (dashed line) with the same
  parameters as those in Fig.\ref{fig:landau}.
           Relativistic effects does not appear in 
the Rayleigh peak as in Landau case, but
more enhance the Brillouin peaks 
than in the Landau case. Note that the scale of the vertical
line is much bigger than that in Fig.~\ref{fig:landau}.
           }
  \label{fig:tko}
 \end{minipage}
\end{figure}%

The resultant spectral function is shown in  
Fig.\ref{fig:tko} together with that of the minimal relativistic case; the parameters are the same 
as in Fig.\ref{fig:landau}:
As in the case of the Landau equation,
any relativistic effects does not appear in the Rayleigh peak 
while the Brillioun peaks are enhanced by them.
The enhancement is more prominent than in the Landau case 
because the bulk viscosity is involved in addition to the thermal
conductivity as relativistic effects.
However, it is noteworthy that
 relativistic effects only appears in the
Brillouin peaks but not in the Rayleigh peak,
irrespective of the choice of the frame of the
fluid dynamic equation.

\subsection{In the case of Israel-Stewart equation in particle frame}

The Israel-Stewart (I-S) equation\cite{is} is a second-order equation where
 relaxation times are contained and has a form of 
the telegrapher's equation.
Thus in the I-S equation,
the so called causality problem is formally resolved.
However, 
the fluid dynamic equations are 
supposed to describes the behaviors of the conserved quantities 
in even larger scale than the mean free path,
and the problem of causality  takes place in short wavelength region.
Therefore,  it is expected that the acausal and
causal fluid dynamic equations should give
the same description for the phenomena with  long wavelengths,
or hydrodynamic modes.
In fact we shall show that the I-S equation gives 
completely the same results for the spectral
function of the density fluctuations
as the Landau equation gives.

In the I-S equation in the particle frame, 
the dissipative terms are given by 
\begin{equation}
\tau^{\mu \nu}=-\Pi\Delta^{\mu \nu}+q^\mu u^\nu +q^\nu u^\mu +\pi^{\mu \nu}, 
\end{equation}
and
$\nu^\mu=0$.
Here 
\beq
\Pi &=& 
-\zeta(\partial_\mu u^\mu +\beta_0 u^a\partial_a \Pi -\alpha_0 \partial_\mu q^\mu),\\ 
q^\mu& =&\kappa T \Delta^{\mu \nu}(\frac{1}{T}\partial_\nu T
       -u^a\partial_a u_\nu 
       -\beta_1 u^a\partial_a q_\nu
       -\alpha_0 \partial_\nu \Pi +\alpha_1 \partial_a \pi^a_\nu  ), \\
\pi^{\mu \nu}&=& 2\eta \Delta^{\mu \nu a b}( \partial_a u_b
              -\beta_2 u^c \partial_c \pi_{a b}
              -\alpha_1 \partial_a q_b  )  ,
\eeq
with 
$u^\mu q_\mu=0,$ \, 
$\pi^{\mu \nu}=\pi^{\nu \mu}$,\, 
$u^\mu \pi_{\mu \nu}=0$ and 
$\pi^\mu_\mu=0$.
Here $\beta_0 $, $\beta_1 $ and $\beta_2 $ are the relaxation time of   
the bulk viscosity, the heat flux and the shear viscosity, respectively.
$\alpha_0$ ($\alpha_1$) is the coupling of the bulk viscosity and the heat flux 
(the shear viscosity and the heat flux).
 $\Delta^{\mu \nu \rho \sigma}$ is a projector defined by   
\begin{equation}
\Delta^{\mu \nu \rho \sigma} =\frac{1}{2}[\Delta^{\mu \rho}\Delta^{\nu \sigma}
                      +\Delta^{\mu \sigma} \Delta^{\nu \rho} 
                      -\frac{2}{3}\Delta^{\mu \nu} \Delta^{\rho \sigma}], 
\end{equation} 
Applying the similar procedure as in the first-order equations,
we find the time-dependent density fluctuation is given by
\beq
  \delta n(\bfk ,t)/\delta n(\bfk ,0) &\sim &  
     (1-\frac{1}{\gamma})
     \e^{-\chi k^{2}t} + \frac{1}{\gamma}
     \cos(c_s k t) \e^{-\Gamma_{\rm B} k^{2} t}\nonumber  \\
    & &+O(k^2) \times \big[\e^{-t/\beta_0 \zeta}+\e^{-t/2\beta_2 \eta}
     + \e^{-w_0 t /[(\beta_1 w_0-1)\kappa T_0]}\big],
   \label{eq:isdensity}
\eeq
with
\begin{equation}
\Gamma_{\rm B}=\frac{1}{2}[(\gamma-1)\chi+\nu_l +c_s^2 T_0(\kappa /w_0-2\chi\alpha_P )]
\equiv \Gamma^{\rm IS}_{\rm B},
\end{equation}
which is exactly the same as $\Gamma^{\rm La}_{\rm B}$ 
given in the Landau equation in the energy frame.
See Appendix D for a detailed derivation of these results.
We note that the last term in Eq.(\ref{eq:isdensity}) 
\begin{equation}
{\rm exp}[-\frac{w_0 t}{(\beta_1 w_0-1)\kappa T_0}],
\label{eq:pc}
\end{equation}
has an exponent which can be positive depending  on the value of the relaxation time
$\beta_1$.
In fact this term is a remnant existing for the
Eckart equation, which as shown in Appendix B, behaves pathologically
as
\begin{equation}
\delta n(\bfk ,t) \sim {\rm exp}[\frac{w_0}{\kappa T_0}t].
\end{equation}
From Eq.(\ref{eq:pc}), one can see that 
if the relaxation time of the heat current is large enough to satisfy
the inequality 
\begin{equation}
\beta_1 > \frac{1}{w_0},
\end{equation}
the density fluctuation would relax to the equilibrium state.
On the other hand, if the relaxation time of the heat current is
so small as to satisfy the condition
\begin{equation}
\beta_1 < \frac{1}{w_0},
\end{equation} 
the density fluctuation will not relax down and the system stays pathological
even though the relaxation time is finite;  
that is, the Israel-Stewart equation in particle frame takes over the pathological 
behavior of Eckart equation. 

Here let us assume that the relaxation time is sufficiently large
to satisfy the inequality $\beta_1 > \frac{1}{w_0}$.
Then  the spectral function can be obtained as 
\begin{eqnarray}
 \frac{S_{n n}(\bfk ,\omega )}{\la (\delta n(\bfk ,t=0))^2\ra }=
  &&(1-\frac{1}{\gamma})\frac{2\chi k^{2}}{\omega^{2}+\chi^{2}k^{4}} 
   +\frac{1}{\gamma}[\frac{\Gamma_{\rm B} k^{2}}{(\omega -c_{s}k)^{2}+\Gamma_{\rm B}^{2}k^{4}} \nonumber \\
     &&+\frac{\Gamma_{\rm B} k^{2}}{(\omega +c_{s}k)^{2}+\Gamma_{\rm B}^{2}k^{4}}]      
     +O(k^2) \times [\frac{2  /\beta_0\zeta}{\omega^{2}+1/(\beta_0\zeta )^2}  \nonumber \\  
       &&+\frac{1  /\beta_2\eta}{\omega^{2}+1/(2\beta_2\eta)^2}
       +\frac{2 w_0 /[(\beta_1 w_0-1)\kappa T_0]}
       {\omega^{2}+w_0^2 /[(\beta_1 w_0-1)\kappa T_0]^2} ].
 \label{eq:is}
\end{eqnarray}
Apparently, the spectral function has six peaks including the conventional
three peaks,
but the new  three Lorentzian functions should vanish in the long wavelength limit 
$k \rightarrow 0$,
because the strength of these is in the second order of $k$.
Therefore Israel-Stewart equation gives completely 
 the same result for the spectral function of the dynamical
fluctuations as Landau equation does
in the long wavelength limit, as shown in Fig.\ref{fig:landau}.
That is, the relaxation times does not affect the result in the 
fluid dynamical regime.

Now one sees that the two relativistic fluid dynamic equations in the particle frame
give different results on the dynamical density fluctuation;
I-S equation gives  the same result as that given by the equation in the energy frame
whereas the TKO equation shows a frame dependence in the spectral peak
of the sound mode.
What is the origin of the difference?
It may be attributed to that of 
the condition for the dissipative part of the energy-momentum tensor
$\delta T^{\mu\nu}$ in these two equations though in the same particle
frame: The I-S equation\cite{is}  is constructed with an ad-hoc postulate
on $\delta T^{\mu\nu}$ as in the Eckart equation\cite{eckart} that 
$u_{\mu}\delta T^{\mu\nu}u_{\nu} =0$ which the 
Landau equation in the {\em energy frame} does satisfy because
 $u_{\mu}\delta T^{\mu\nu}=0$ by the very definition of the energy frame
by Landau and Lifshitz\cite{landau}.
On the other hand, the TKO equation\cite{tko} in the particle frame is derived by a
powerful reduction theory called the renormalization-group(RG) method
\cite{rg} from the relativistic Boltzmann equation and
it is found\cite{tko,tk} that  their equation satisfies $\delta T^{\mu}_{\mu}=0$
but not $u_{\mu}\delta T^{\mu\nu}u_{\nu}=0$. It is to be noted that
the RG method gives the Landau equation 
for the fluid dynamic equation in the energy frame\cite{tko}. 
It is clear that more work is needed to establish the correct fluid dynamic
equation in the particle frame.

\section{The behavior around the QCD critical point}

We analyze the behavior of the spectral function of the density fluctuations
around the QCD CP on the basis of the dynamic as well as static
scaling laws\cite{stanley,onuki}.
We introduce
the static critical exponents $\tilde{\gamma}$ and $\tilde{\alpha}$ which are defined as follows 
\begin{equation}
\tilde{c}_n = c_0 t^{-\tilde{\alpha}}, \hspace{0.5cm} K_T = K_0 t^{-\tilde{\gamma}},
\end{equation}
where $t=\vert (T - T_c) / T_c \vert$ is a reduced temperature,
$c_0$ and $K_0$ are constants and $K_T=(1/n_0)(\partial n/\partial P)_T$ is the isothermal compressibility.
Using the known thermodynamical identities\cite{stanley} 
\begin{eqnarray}
\tilde{c}_P &=& \tilde{c}_n+(T_0/n_0)\bl( \frac{\partial P}{\partial T} \br)_n^2 K_T, \\
c_s^2         &=& \frac{\tilde{c}_P}{w_0 \tilde{c}_n K_T }, \\
\alpha_P    &=& \bl( \frac{\partial P}{\partial T} \br)_n K_T, 
\end{eqnarray}
we see that the specific heat at constant pressure, the sound velocity 
and the thermal expansivity at constant pressure has the critical behavior as  
\begin{eqnarray}
\tilde{c}_P &\sim & \frac{K_0 T_0}{n_0} \br( \frac{\partial P}{\partial T } \bl)_n^2 t^{-\tilde{\gamma}}, \\
c_s^2       &\sim & \frac{T_0}{n_0 w_0 c_0} \br( \frac{\partial P}{\partial T}\bl)_n^2 t^{\tilde{\alpha}},\\
\alpha_P    &\sim & K_0\br( \frac{\partial P}{\partial T}\bl)_n t^{-\tilde{\gamma}},      
\end{eqnarray}
respectively.
Since the phase transition at the QCD CP belongs 
to the same universality class Z$_2$ as the liquid-gas CP, 
the critical exponents are given as 
$\tilde{\alpha} \sim  0.11$ and 
$\tilde{\gamma} \sim  1.2$.
Recently, it has been argued \cite{Moore} on the basis of the analysis 
of the liquid-gas CP by Onuki\cite{onuki:1997} that 
the bulk viscosity may show a singular behavior around the QCD CP.
So we introduce the critical exponent $a_{\zeta}$ as
\begin{equation}
\zeta = \zeta_0 t^{-a_{\zeta}},
\end{equation}
where $\zeta_0$ is constant.
The exponent $a_{\zeta}$ for the liquid-gas CP
is predicted to be $z\nu-\tilde{\alpha}$ \cite{onuki:1997}.
Here $z$ is the dynamical critical exponent 
and $\nu$ is the exponent which represents the singularity of the correlation length.
These are given by $z \sim 3$  and $\nu \sim 0.63$ 
around the liquid-gas CP.

We denote the exponent of the thermal conductivity by 
$a_{\kappa}$,
\begin{equation}
\kappa =\kappa_0 t^{-a_{\kappa}},
\end{equation}
where $\kappa_0$ is a constant.
It is known that $a_{\kappa} \sim 0.63$ around the liquid-gas CP.
Let us take it for granted that the exponents $a_{\zeta}$ and $a_{\kappa}$ at the 
QCD CP are
given by those at the liquid-gas CP.
That is, we take
\begin{eqnarray}
a_{\zeta} &&\sim z\nu-\tilde{\alpha} \sim 1.8, \\
a_{\kappa} && \sim 0.63.
\end{eqnarray}
Combining these ingredients including the possible singular behavior of $\zeta$,  
we see that the width of the Rayleigh peak behaves as
\begin{equation}
\Gamma_{\rm R} \sim \frac{\kappa_0}{K_0 T_0} 
\br( \frac{\partial T}{\partial P}\bl)_n^2 t^{\tilde{\gamma}-a_{\kappa}}.
\end{equation}
Then, we see
\begin{equation}
\Gamma_{\rm R}  \sim t^{\tilde{\gamma}-a_{\kappa}},
\label{eq:gr}
\end{equation} 
which shows that the width $\Gamma_{\rm R}$ becomes narrow as the QCD CP is approached.   
We emphasize that this result is independent of the choice of the 
relativistic fluid dynamic equation or frame.

So far for the width of the Rayleigh peak.
How about the width $\Gamma_{\rm B}$ of the Brillouin peaks?
For the Landau and I-S cases, 
it has the critical behavior as follows,
\begin{equation} 
\Gamma_{\rm B} \sim \frac{\zeta_0}{2 w_0} t^{-a_{\zeta}}.
\label{eq:gb}
\end{equation}
We note that this singularity comes from that of the bulk viscosity.    

In the case of TKO equation, we first note that 
the critical behavior of the effective bulk viscosity is given by
\begin{equation}
\zeta^{'} = \frac{\zeta}{(3\gamma -4)^2} \sim t^{2\tilde{\gamma}-a_{\zeta}},
\end{equation} 
which shows that the effective bulk viscosity has a positive exponent and 
does not show a singular behavior
because $a_{\zeta} \sim 1.8$ and $\tilde{\gamma} \sim  1.2$.
Instead, the singularity of the Brillouin peaks for TKO equation 
comes from that of the thermal conductivity; 
\begin{equation} 
\Gamma_{\rm B} \sim \Gamma_{\kappa}^{\rm TKO} t^{-(a_{\kappa}-\tilde{\alpha})},
\end{equation}
with
\begin{equation}
\Gamma_{\kappa}^{\rm TKO} = \frac{\kappa_0}{2 c_0 n_0}
                    \bl[ 1-2\frac{T_0}{w_0}\bl( \frac{\partial P}{\partial T}\br)_n \br] .                          
\end{equation}
We note that the strength of the divergence of $\Gamma_{\rm B}$ for TKO equation is
weaker  than that for the Landau and I-S equations.

Anyway, we have confirmed that the width $\Gamma_{\rm B}$ may diverge at the QCD CP 
irrespective of the relativistic fluid dynamic equations.

\begin{figure}[!t]
  \centering
   \includegraphics[width=70mm, angle=270]{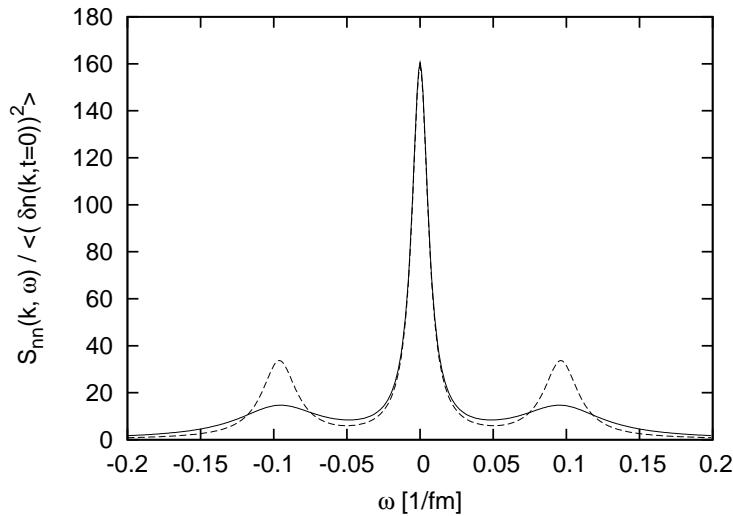}
  \caption{
The spectral function at $t=0.5$ and $k=0.1$\, [1/fm].
           The solid line represents the Landau/Israel-Stewart case, while
           the dashed line  the TKO case. 
           The strength of the Brillouin peaks becomes small due to the singularity of
 the ratio of specific heats.
 }
  \label{fig:t5}
\end{figure}
\begin{figure}[!t]
  \centering
   \includegraphics[width=70mm, angle=270]{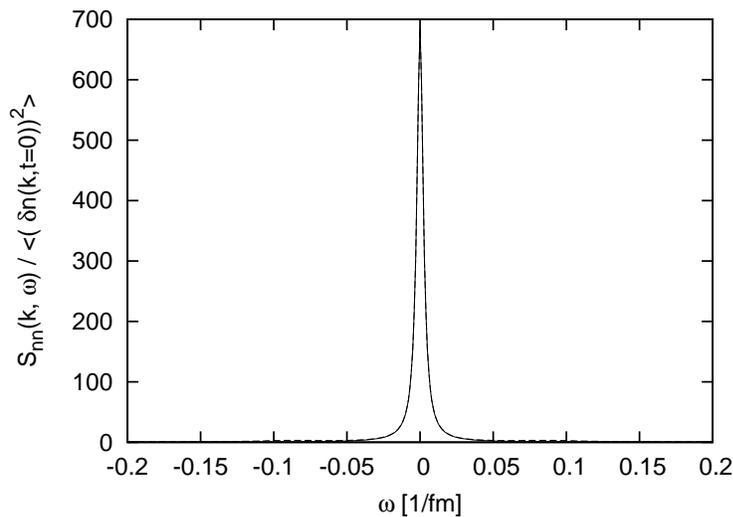}
  \caption{ 
The spectral function at $t=0.1$ and $k=0.1$\, [1/fm].
           We see that the Brillouin peaks which correspond to sound wave dies out
           and the difference between the Landau and TKO cases disappears.
Note that the scale of the vertical line is much bigger than that of
Fig.\ref{fig:t5}.
}
  \label{fig:t1}
\end{figure}

Unfortunately or fortunately,
these singular behaviors of the width of the Brillouin peaks  around the
 QCD CP may not be observed:
Note that the strengths of the Rayleigh and the Brillouin peaks
are given in terms of $\gamma$, the ratio of the specific heats, which
 behaves like 
\begin{equation}
\gamma = \tilde{c}_p / \tilde{c}_n \sim t^{-\tilde{\gamma}+\tilde{\alpha}}
\rightarrow \infty,
\end{equation}
in the critical region. 
Then the strength of the Brillouin peaks is attenuated 
and only the Rayleigh peak stands out in the critical region, as follows;
\begin{eqnarray}
S_{n n}(\bfk ,\omega ) 
\sim  \la (\delta n(\bfk ,t=0))^2\ra \frac{2\Gamma_{\rm R} k^{2}}{\omega^{2}+\Gamma_{\rm R}^{2}k^{4}},
\quad (t\sim 0).
\end{eqnarray}
 Figure~\ref{fig:t5} and \ref{fig:t1} show
 how the spectral function behaves around the QCD CP
for $k=0.1$\, [1/fm],\, $t=0.5$ and $t=0.1$, respectively.
We see that the strength of the Brillouin peaks becomes small and dies out as the system approach the QCD CP. 
In addition, the static  correlation function $\la (\delta n(\bfk ,t=0))^2\ra$ has the critical behavior
as
\begin{equation}
\la (\delta n(\bfk ,t=0))^2\ra \sim \frac{B}{1+\xi^2 k^2 },
\label{eq:co} 
\end{equation}
with $B=T K_T / V$, where V is the volume\cite{Gebhardt}.
Equation (\ref{eq:co}) would show a singular peak in the forward direction $k=0$,
because $K_T\sim t^{-1.2}$ and $\xi^2\sim t^{-1.26}$.
\footnote{This is known as the critical opalescence\cite{stanley,reichl}.}
Then the strength of the Rayleigh peak will be
most  drastically enhanced in the forward angle in the critical region.

The critical behavior on the spectral function of the density fluctuation 
is summarized as follows.
\begin{enumerate}
\item Disappearance of the strength of the sound mode:\\
The sound mode loses its strength at the QCD CP
because of the divergence of the ratio of the specific heats.
Thus, we see that the formation of the sound mode itself is suppressed. 

\item Divergence of the width of the sound mode $\Gamma_{\rm B}$:\\
According to Eq.(\ref{eq:gb}), the width of the sound mode 
 diverges at the QCD CP,
which means that the sound mode is strongly damped and attenuated in the critical region. 
Therefore we see that the sound mode is very much attenuated 
even if it is created with a small strength.

\item Narrowing of the width of the thermal mode $\Gamma_{\rm R}$:\\
The width of the thermal mode vanishes at the QCD CP 
because of the divergence of the specific heat at constant pressure.
This means that the entropy density exhibits critical slowing down and 
the thermal mode is the critical mode at the QCD CP.
That is, the entropy density is the order parameter and mainly governs 
the critical dynamics at the QCD CP as in the liquid-gas CP.
\end{enumerate} 

\section{Discussions}

We have seen that the sound mode would lose its strength and be strongly damped as the system
approaches the critical point.
But why at all do sound modes die out at the critical point? 
To answer this question, let $\xi=\xi_0t^{-\nu}$ be the correlation length which diverges
as the critical point is approached. 
If we write the wave length of the sound mode by $\lambda_s$,
the fluid dynamic regime is expressed as
\beq
\xi << \lambda_s,
\eeq
with which condition the sound mode can develop\cite{HALPERIN:1969zza}. 
However, in the vicinity of the critical point, the correlation length
$\xi$ becomes very large and eventually goes to infinity, so
the above inequality can not be satisfied, and the sound mode
can not be developed in the vicinity of the critical point\cite{HALPERIN:1969zza}.

From this argument, we can speculate that  phenomena inherently related to the existence
of the sound mode may disappear around the critical point.
One of such phenomena is the possible Mach cone formation \cite{Torrieri:2009mv} 
by the particle passing through 
the medium with a speed larger than the sound velocity $c_s$.
Such a Mach-cone like particle correlations 
are observed in the RHIC experiment\cite{:2008nd}. 
If such three-particle correlations have been 
confirmed to be a Mach-cone formation,
then the disappearance or suppression of
the Mach cone would be a signal that the created matter 
has passed through the critical region, showing the existence of
the QCD critical point. Even if the thermal wake also contributes
to the formation of Mach cone\cite{betz}, a suppression of Mach cone may be 
expected by the attenuation of the sound mode. 
So it would be very interesting to see possible
variation of the strength of the Mach cone 
according to the variation of the incident energy of the heavy-ion collisions.
In theoretical side, it is an intriguing task to explore 
the fate of Mach cone with an EoS which admits the existence of the CP.

Precisely speaking,
the analysis based on  fluid dynamics is, however, 
only valid in the fluid dynamical regime: $k\xi << 1$,
that is, our discussions on the critical behavior are
based on  extrapolation from the fluid dynamical to the 
critical region: $k\xi >> 1$.\cite{HALPERIN:1969zza}
This extrapolation is known to make good sense 
for the non-relativistic case\cite{stanley}. 
It is well known that the  coupling of the thermal mode  to 
the diffusive transverse mode becomes essential for the description of the
dynamical critical phenomena in the close vicinity of the CP 
 for the non-relativistic case\cite{kawasaki}.  
This coupling can be analyzed by the mode-mode coupling theory\cite{kawasaki,onuki}.
Such an analysis is interesting but beyond the scope of the present work.

\section{Summary and concluding remarks} 

We have explored how the singularities of 
the thermodynamic values as well as the transport coefficients affect the 
dynamical density
 fluctuations around the QCD critical point(CP) using dissipative relativistic fluid dynamics.
Our analysis is an extension of those made for non-relativistic 
case to the relativistic case. 
We have shown that  the sound modes which are  directly related to the density fluctuation
are enhanced by relativistic effects, but 
tend to be attenuated  around and would eventually
die out at the CP and, hence,  are not the soft mode.
Our analysis based on the relativistic fluid dynamics has shown that 
the genuine and remaining soft mode at the QCD CP is the diffusive thermal mode.
Our analysis also suggests that the possible divergent behavior of the bulk viscosity 
may not
be observed through the density fluctuations.
We have also argued 
that the possible suppression or disappearance of Mach cone 
can be used as a signal of the existence of the QCD CP.

We have shown that the relativistic effect on the spectral function 
of the density fluctuation appears only in the width of the Lorentzian function.
More precisely,
the relativistic effect enhances only the Brillouin peaks due to the sound mode
but not the Rayleigh peak irrespectively of the equations
and the frames.
Therefore one might tend to think that 
the choice of the frame is just a matter of convenience 
and gives no physical difference. 
However, it is noteworthy that there is a slight difference in the enhancement
depending on the choice of the fluid dynamic equation in the particle frame.
The spectral function obtained in the Israel-Stewart(I-S) equation gives the
same result as that in the Landau equation in the energy frame whereas
the equation by Tsumura-Kunihiro-Ohnishi(TKO)\cite{tko} gives a  more
enhanced peaks to the sound mode than in the two equations.
We have suggested that the origin of the difference may be attributed to that of 
the condition for the dissipative part of the energy-momentum tensor
$\delta T^{\mu\nu}$ in these two equations though in the same particle
frame.
It is certain that more work is needed to establish the correct fluid dynamic
equation in the particle frame\cite{tk2}.

We have also elucidated the  properties of some of dissipative relativistic
fluid dynamic equations:
In the case of Eckart equation, the density fluctuation around the
equilibrium is unstable and tends to  diverge as $t$ goes infinity.
This pathological behavior is due to the heat current induced by the time 
derivative of the fluid velocity.    
Since this property is carried over to the Israel-Stewart(I-S) equation in particle frame,
the density fluctuation in the I-S equation 
also behaves pathologically if the relaxation time of the system
happens to be small,  
although it {\em is} formally causal;
if the relaxation time is large enough,
the I-S equation in particle frame gives completely the same  spectral function
for the density fluctuations as Landau equation gives.
That is, formally acausal fluid dynamic equations suffices in description of
phenomena in the fluid dynamical regime with long wave lengths
 if the equation gives a relaxation of the fluctuations around the thermal
equilibrium.

In the present work, we have confined ourselves to 
the study of fluctuations around the  thermal equilibrium.
To make a direct relevance to the relativistic heavy-ion collisions,
 study on the fluctuations around expanding flow background like Bjorken flow 
should be done, which constitute one of the  future works.
It should be also an intriguing task to
explore the fate of Mach cones or shock waves in general 
in the critical region by explicit calculations.

It is important to study the coupling between 
the thermal fluctuations and the transverse mode
using the mode-mode coupling theory 
in the close vicinity of the critical region for the relativistic case.
The present work constitutes  the basis for such a more complete analysis.

\section*{Acknowledgments}
We would like to thank Kenji Fukushima, Hideo Suganuma,
Akira Ohnishi and the members of the Quark-Hadron seminar at
Kyoto university for their interest in our work and encouragement.
Discussions during the YIPQS international molecule-type workshop 
on "Non-equilibrium quantum field theories and dynamic critical phenomena", 
March 2009, were useful to complete this work. 
We  acknowledge the participants of the workshop, especially Juergen Berges,
 Hirotsugu Fujii, Berndt Mueller and Misha Stephanov for their
interest in this work and comments.
We thank Guy Moore for giving our attention to 
Ref.\citen{Moore}.
This work was partially supported by a
Grant-in-Aid for Scientific Research by the Ministry of Education,
Culture, Sports, Science and Technology (MEXT) of Japan (Nos.
20540265, 19$\cdot$07797),
 by Yukawa International Program for Quark-Hadron Sciences, and by the
Grant-in-Aid for the global COE program `` The Next Generation of
Physics, Spun from Universality and Emergence '' from MEXT.


\appendix
\section{derivation of thermodynamic identities}
In this Appendix, we derive thermodynamic identities we have used.
Here $s$ is not the entropy density but the entropy per the particle number,
$\tilde{c}_n=T_0(\partial s / \partial T)_n$ and $\tilde{c}_P=T_0(\partial s / \partial T)_P$
are the specific heats at constant density and pressure, respectively, 
$c_s=(\partial P/\partial \epsilon)_s^{1/2}$ is the sound velocity, 
$\alpha_P=-(1/n_0)(\partial n / \partial T)_P$ is the thermal expansivity 
at constant pressure, and $\gamma=\tilde{c}_P/\tilde{c}_n$ is the ratio of the specific 
heats.
$\frac{\partial(a,b)}{\partial(c,d)}$ represents Jacobian determinant: 
\begin{equation}
\frac{\partial(a,b)}{\partial(c,d)} \equiv 
 \frac{\partial a}{\partial c}\frac{\partial b}{\partial d}
 -\frac{\partial a}{\partial d}\frac{\partial b}{\partial c}.
\end{equation}

\subsection{Derivation of $(\partial P / \partial n)_{T}=(w_{0}c_{s}^{2})/(n_{0}
\gamma)$}

\begin{eqnarray}
\bl(\frac{\partial P}{\partial n}\br)_{T}&=& \frac{\partial(P,T)}{\partial(n,T)}, \nonumber \\
 &=& \frac{\partial(P,T)}{\partial(P,s)}
     \frac{\partial(P,s)}{\partial(\epsilon,s)}
     \frac{\partial(\epsilon,s)}{\partial(n,s)} 
     \frac{\partial(n,s)}{\partial(n,T)},                                   \nonumber \\
 &=& \bl(\frac{\partial T}{\partial s}\br)_{P}
     \bl(\frac{\partial P}{\partial \epsilon}\br)_{s}
     \bl(\frac{\partial \epsilon}{\partial n}\br)_{s}
     \bl(\frac{\partial s}{\partial T}\br)_{n},                                   \nonumber \\
 &=& \frac{T}{\tilde{c}_{P}} \cdot
     c_{s}^{2} \cdot 
     \bl(\frac{\partial \epsilon}{\partial n}\br)_{s} \cdot
     \frac{\tilde{c}_{n}}{T},                                               \nonumber \\
 &=& \frac{c_{s}^{2}}{\gamma}\bl(\frac{\partial \epsilon}{\partial n}\br)_{s},
\end{eqnarray}
where
\begin{eqnarray}
\bl(\frac{\partial \epsilon}{\partial n}\br)_{s} 
&=& \frac{\partial(\epsilon, s)}{\partial (n, s)},  \nonumber \\
&=& \frac{\partial(\epsilon, s)}{\partial (\epsilon, n)}
    \frac{\partial(\epsilon, n)}{\partial (n, s)},  \nonumber \\
&=& -\bl(\frac{\partial s}{\partial n}\br)_{\epsilon}
    \bl(\frac{\partial \epsilon}{\partial s}\br)_{n}.
\end{eqnarray}
Using the first law of thermodynamics
\begin{equation}
d\epsilon = T d(n s)+\mu d n,
\end{equation}
we find 
\begin{equation}
\bl(\frac{\partial \epsilon}{\partial(n s)}\br)_{n}=T
\rightarrow
\bl(\frac{\partial \epsilon}{\partial s}\br)_{n}=n T,
\end{equation}
and 
\begin{eqnarray}
\bl(\frac{\partial (n s)}{\partial n}\br)_{\epsilon} &=& -\frac{\mu}{T}, \nonumber \\
s+n\bl(\frac{\partial s}{\partial n}\br)_{\epsilon} &=& -\frac{\mu}{T}, \nonumber \\
\bl(\frac{\partial s}{\partial n}\br)_{\epsilon}    &=& -\frac{Ans+n\mu}{n^{2}T}
=-\frac{w}{n^{2}T},
\end{eqnarray}
where we have used $\epsilon=Tns-P+\mu n $ and $w=\epsilon+P$.\\
Then we obtain
\begin{equation}
\bl(\frac{\partial \epsilon}{\partial n}\br)_{s}=\frac{w}{n},
\end{equation}
and so 
\begin{equation}
\bl(\frac{\partial P}{\partial n}\br)_{T}=\frac{w_{0}c_{s}^{2}}{n_{0}\gamma},
\label{eq:a}
\end{equation}

\subsection{Derivation of $(\partial P / \partial T)_{n}=
w_{0}\alpha_{P}c_{s}^{2}/\gamma $}
\begin{eqnarray}
\bl(\frac{\partial P}{\partial T}\br)_{n}&=& \frac{\partial (P, n)}{\partial (T, n)}, \nonumber \\
 &=& \frac{\partial (P, n)}{\partial (T, P)}
     \frac{\partial (T, P)}{\partial (s, P)}
     \frac{\partial (s, P)}{\partial (s, \epsilon)}
     \frac{\partial (s, \epsilon)}{\partial (s, n)}
     \frac{\partial (s, n)}{\partial (T, n)},  \nonumber \\
 &=& -\bl(\frac{\partial n}{\partial T}\br)_{P}
      \bl(\frac{\partial T}{\partial s}\br)_{P}
      \bl(\frac{\partial P}{\partial \epsilon}\br)_{s}
      \bl(\frac{\partial \epsilon}{\partial n}\br)_{s}
      \bl(\frac{\partial s}{\partial T}\br)_{n},     \nonumber \\
 &=& n_{0} \alpha_{P} \cdot
     \frac{T}{\tilde{c}_{P}}  \cdot
     c_{s}^{2}  \cdot
     \frac{w_{0}}{n_{0}}  \cdot
     \frac{\tilde{c}_{n}}{T},                  \nonumber \\
 &=& \frac{w_{0}\alpha_{P}c_{s}^{2}}{\gamma}.
\end{eqnarray}

\subsection{Derivation of$(\partial s / \partial n)_{T}=-(w_{0}
\alpha_{P}c_{s}^{2})/(n_{0}^{2}\gamma) $}
Assuming that the total particle number N is constant, we find  
\begin{eqnarray}
\bl(\frac{\partial s}{\partial n}\br)_{T}
&=&\bl(\frac{\partial (S/N)}{\partial (N/V)}\br)_{T}, \nonumber \\
&=&\frac{1}{N^{2}}\frac{\partial V}{\partial (1/V)}
\bl(\frac{\partial S}{\partial V}\br)_{T}, \nonumber \\
&=&-\frac{1}{n_{0}^{2}}\bl(\frac{\partial S}{\partial V}\br)_{T}.
\end{eqnarray}
Using Maxwell relations 
\begin{equation}
\bl(\frac{\partial S}{\partial V}\br)_{T}=\bl(\frac{\partial P}{\partial T}\br)_{V}.
\end{equation}
We obtain
\begin{eqnarray}
\bl(\frac{\partial s}{\partial n}\br)_{T}
&=&-\frac{1}{n_{0}^{2}}\bl(\frac{\partial P}{\partial T}\br)_{V}, \nonumber \\
&=&-\frac{1}{n_{0}^{2}}\bl(\frac{\partial P}{\partial T}\br)_{n}, \nonumber \\
\bl(\frac{\partial s}{\partial n}\br)_{T}&=&-\frac{w_{0}\alpha_{P}c_{s}^{2}}{n_{0}^{2}\gamma}.
\end{eqnarray}
where we have used Eq.(\ref{eq:a}).

\subsection{Derivation of $(w_{0}T_{0}\alpha_{P}^{2}c_{s}^{2})/(n_{0}\tilde{c}_{P}\gamma)
=1-1/\gamma $}
\begin{eqnarray}
\tilde{c}_{P} &=& T_{0}\frac{\partial (s,P)}{\partial (T,P)}, 
                                     \nonumber \\
              &=& T_{0}\frac{\partial (s,P)}{\partial (T,n)}\frac{\partial (T,n)}{\partial 
(T,P)}, \nonumber \\
              &=& T_{0}\bl(\frac{\partial s}{\partial T}\br)_{n}
                 +T_{0}\bl(\frac{\partial s}{\partial n}\br)_{T}\bl(\frac{\partial n}{\partial T}\br)_{P},
     \nonumber \\
\tilde{c}_{P}-\tilde{c}_{n} 
              &=& T_{0}\bl(\frac{\partial s}{\partial n}\br)_{T}\bl(\frac{\partial n}{\partial T}\br)_{P},
   \nonumber \\              
              &=& T_{0} \cdot \bl(-\frac{w_{0}\alpha_{P}c_{s}^{2}}{n_{0}^{2}\gamma} \br)
                  \cdot (- n_{0} \alpha_{P}).
\end{eqnarray}
Dividing the both side by $\tilde{c}_{P}$, we obtain
\begin{equation}
1-\frac{1}{\gamma}=\frac{w_{0}T_{0}\alpha_{P}^{2}c_{s}^{2}}{n_{0}\tilde{c}_{P}\gamma}.
\label{eq:tid}
\end{equation}

\section{In the case of Eckart equation (particle frame)}

In this Appendix, we shall show explicitly how the pathological behavior
of the Eckart equation
hinders the procedure of the calculation of the spectral function 
of the density fluctuations for an instructive purpose.

In  Eckart equation, the dissipative terms are given by
\begin{eqnarray}
\tau^{\mu\nu}=&\eta&[\,\partial^{\mu}_{\perp}u^{\nu}+
   \partial^{\nu}_{\perp}u^{\mu}-\frac{2}{3}\Delta^{\mu\nu}(\partial_{\perp}{\cdot}u)\,]
   +\zeta\Delta^{\mu\nu}(\partial_{\perp}{\cdot}u), \nonumber \\
   &+&\kappa(u^{\mu}\partial^{\nu}_{\perp}T+u^{\nu}\partial _{\perp}^{\mu}T
   -T u^\mu (u \cdot \partial )u^\nu -T u^\nu (u \cdot \partial )u^\mu ), 
   \label{eq:eck}\\
\nu^{\mu}=&0&.
\end{eqnarray}
The term $-T u^\mu (u \cdot \partial )u^\nu -T u^\nu (u \cdot \partial )u^\mu$ in 
Eq.(\ref{eq:eck})
represents the heat current induced by the time derivative of the fluid velocity,
and this term is identified to the origin 
of the pathological behavior.   

The linearized fluid dynamic equation reads 
\begin{eqnarray} 
\frac{\partial \delta n}{\partial t}+n_{0}\nabla\cdot\delta\bfv &&=0, \\
  w_{0}\frac{\partial \delta\bfv }{\partial t}
  -\eta\nabla^{2}\delta\bfv 
  -(\frac{1}{3}\eta+\zeta^{'})\nabla(\nabla\cdot\delta\bfv )+\nabla(\delta P)
  -\kappa\nabla&&(\frac{\partial \delta T}{\partial t}) \\
  -\kappa T_0 \frac{\partial^2 \delta\bfv }{\partial t^2}&&=0, 
  \label{eq:eckart}     \\
n_{0} \frac{\partial \delta s}{\partial t}
  -\kappa \nabla \cdot \frac{\partial \delta\bfv }{\partial t}
  -\frac{\kappa}{T_{0}}\nabla^{2}\delta T&&=0.
\end{eqnarray}
It is noted that the second order derivative in time appear in Eq.(\ref{eq:eckart})
which arises from the heat current induced by the time derivative of 
the fluid velocity. 
From the same procedure as was done for the Landau equation in $\S$ 2, we obtain
\begin{equation}
A  
  \begin{pmatrix}
  \delta n(\bfk ,z) \\
  \delta v_{\parallel}(\bfk ,z) \\
  \delta T(\bfk ,z)
  \end{pmatrix}
  =  
  \begin{pmatrix}
  \delta n(\bfk ,0) \\
  -\kappa \frac{T_0}{w_0}\frac{\partial \delta v_{\parallel}}{\partial t}(\bfk ,0)
  +(-z\frac{\kappa T_0}{w_0}+1)\delta v_{\parallel}
  -i k\frac{\kappa}{w_0}\delta T(\bfk ,0) \\
  -\frac{\alpha_{P}c_{s}^{2}}{n_{0}\gamma}\delta n(\bfk ,0)
  -i k\frac{\kappa}{w_0}\delta v_{\parallel}(\bfk ,0)
  +\frac{n_{0}\tilde{c}_{n}}{T_{0}w_{0}}\delta T(\bfk ,0)
  \end{pmatrix},
\end{equation}
where 
\begin{equation}
  A =
  \begin{pmatrix}
  z  &  i k n_0 & 0  \\
 i k \frac{c_{s}^{2}}{w_{0}\gamma} 
  & -z^2\frac{\kappa T_0}{w_0}+\nu_{l}k^{2} 
  & i k(-\frac{\kappa}{w_0}z+\frac{\alpha_P c_s^2}{\gamma})\\
  -z\frac{\alpha_P c_s^2}{n_0\gamma}
  & -i z k\frac{\kappa}{w_0} 
  & \frac{\tilde{c}_n n_0}{T_0 w_0}(z+\gamma \chi k^2)                                 
  \end{pmatrix}.
\end{equation} 
The Fourier-Laplace transform of the density fluctuation is given by
\begin{equation}
 \delta n(\bfk ,z)=\{ (A^{-1})_{1 1} 
 -\frac{\alpha_{P}c_{s}^{2}}{n_{0}\gamma}(A^{-1})_{1 3}\} \delta n(\bfk ,0),
\end{equation} 
where the irrelevant terms which do not contribute to the density
fluctuations are omitted.
 $\det A$ is evaluated to be 
\begin{eqnarray}
\det A = \frac{\tilde{c}_n n_0}{T_0 w_0}
[-z^4\frac{\kappa T_0}{w_0} 
  +z^3 \{ 1-k^2(\gamma \chi \frac{\kappa T_0}{w_0}+\frac{\kappa^2 T_0}{n_0 w_0
 \tilde{c}_n}) \},  \nonumber \\
  +z^2 k^2(\gamma \chi+\nu_l -2\chi\alpha_p c_s^2 T_0 ) 
  +z(k^2 c_s^2 + k^4 \gamma \chi \nu_l )
  +k^4 \chi c_s^2].
\end{eqnarray}
The fourth order term in $z$ appears in $\det A$
because of the heat current induced by the time derivative of the fluid velocity.
We can factorize this to second order in $k$ 
\begin{equation}
\det A \sim -\frac{\kappa \tilde{c}_n n_0 }{w_0^2}
(z-\frac{w_0}{\kappa T_0})(z+\chi k^2)
(z+\Gamma_{\rm B} k^2+i c_s k)(z+\Gamma_{\rm B} k^2 -i c_s k), 
\label{eq:det}
\end{equation}
where 
\begin{equation}
\Gamma_{\rm B}=\frac{1}{2}[(\gamma -1) \chi +\nu_l +c_s^2 T_0(\kappa/w_0 
-2\chi \alpha_P)  ]
\equiv\Gamma_{\rm B}^{\rm Ec}.
\end{equation}
Comparing Eq.(\ref{eq:det}) with the case of Landau equation,
we see a new mode is present in the present case, which is found to be
an unstable mode so that 
 the density fluctuation increases with $t$ without limit,
\begin{equation}
\delta n(\bfk ,t) \sim exp[\frac{w_0}{\kappa T_0}t].
\end{equation}
Of course, this is an unwanted result
because any fluctuation around the equilibrium state
should relax down to recover the equilibrium state.
This pathological properties of the Eckart equation
was first noted by Hiscock and Lindblom \cite{hiscock}.

\section{Detailed derivation in TKO equation in particle frame}

The linearized TKO equation\cite{tko} reads
\begin{eqnarray}
\frac{\partial \delta n}{\partial t}+n_{0}\nabla\cdot\delta\bfv =0,  \\
w_{0}\frac{\partial \delta\bfv }{\partial t}-\eta\nabla^{2}\delta\bfv 
  -(\frac{1}{3}\eta+\zeta^{'})\nabla(\nabla\cdot\delta\bfv )+\nabla(\delta P)
  -\kappa\nabla(\frac{\partial \delta T}{\partial t})=0, \\
n_{0} \frac{\partial \delta s}{\partial t}
  -\frac{3\zeta^{'}}{T_{0}}\nabla \cdot \frac{\partial \delta\bfv }{\partial t}
  -\frac{\kappa}{T_{0}}\nabla^{2}\delta T=0.
\end{eqnarray}
Accordingly, we have
\begin{equation}
A
  \begin{pmatrix}
  \delta n(\bfk ,z) \\
  \delta v_{\parallel}(\bfk ,z) \\
  \delta T(\bfk ,z)
  \end{pmatrix}
  =  
  \begin{pmatrix}
  \delta n(\bfk ,0) \\
  \delta v_{\parallel}(\bfk ,0)-i k\frac{\kappa}{w_{0}}\delta T(\bfk ,0) \\
  -\frac{\alpha_{P}c_{s}^{2}}{n_{0}\gamma}\delta n(\bfk ,0)
  -i k\frac{3\zeta^{'}}{T_{0}w_{0}}\delta v_{\parallel}(\bfk ,0)
  +\frac{n_{0}\tilde{c}_{n}}{T_{0}w_{0}}\delta T(\bfk ,0)
  \end{pmatrix},
\end{equation}
where
\begin{equation}
A =
  \begin{pmatrix}
  z & i k n_{0} & 0\\
 i k \frac{c_{s}^{2}}{n_{0}\gamma} 
  & z+\nu_{l}k^{2} 
  & i k(-\frac{\kappa}{w_{0}}z+\frac{\alpha_{P} c_{s}^{2}}{\gamma})\\
 -z\frac{\alpha_{P}c_{s}^{2}}{n_{0}\gamma} 
  & -i z k\frac{3\zeta^{'}}{w_{0}T_{0}}
  & \frac{\tilde{c}_{n}n_{0}}{T_{0}w_{0}}(z+\gamma \chi k^{2})
 \end{pmatrix}.
\end{equation}
Using these result, we have the density fluctuation up to the second order in $k$, 
as follows,
\begin{eqnarray}
 \delta n(\bfk ,z)/\delta n(\bfk ,0)
 &=&[(A^{-1})_{1 1}-\frac{\alpha_{P}c_{s}^{2}}{n_{0}\gamma}(A^{-1})_{1 3} ], \nonumber \\
  &\sim &\frac{ (z+\Gamma_{\rm B} k^{2}+i c_{s}k)(z+\Gamma_{\rm B} k^{2}- i c_{s}k)
    -\frac{1}{\gamma}c_{s}^{2}k^{2}+z \chi k^{2}}
    {(z+\chi k^{2})(z+\Gamma_{\rm B} k^{2}+Ac_{s}k)(z+\Gamma_{\rm B} k^{2}-i c_{s}k) }.
\end{eqnarray}
Here
\begin{equation}
\Gamma_{\rm B}=\frac{1}{2}[\chi(\gamma-1)+\nu_{l}^{\rm TKO}
  -\frac{\alpha_{P}c_{s}^{2}}{n_{0}\tilde{c}_{P}}
(\kappa T_{0}+3\zeta^{'})] \equiv \Gamma_{\rm B}^{TKO},
\end{equation}
with 
\begin{equation}
\nu_l^{\rm TKO} = (\zeta^{'}+\frac{4}{3}\eta )/w_0,
\end{equation}
Thus the spectral function is now given by
\begin{eqnarray}
\frac{S_{n n}(\bfk ,\omega )}{\la (\delta n(\bfk ,t=0))^2\ra }=
(1&-&\frac{1}{\gamma})\frac{2\chi k^{2}}{\omega^{2}+\chi^{2}k^{4}}, 
\nonumber \\
&+&\frac{1}{\gamma}
[\frac{\Gamma_{\rm B} k^{2}}{(\omega -c_{s}k)^{2}+\Gamma_{\rm B}^{2}k^{4}}
+\frac{\Gamma_{\rm B} k^{2}}{(\omega +c_{s}k)^{2}+\Gamma_{\rm B}^{2}k^{4}}].
\end{eqnarray}

\section{Detailed derivation in the case of I-S equation}

Applying the similar procedure as in the first-order equations,
the linearized I-S equation reads;
\beq
\frac{\partial \delta n}{\partial t} + n_0 \nabla \cdot \delta \bfv &=&0, \\
(\epsilon_0 + P_0)\frac{\partial \delta \bfv }{\partial t}
          +\nabla(\delta P+\delta \Pi) +\frac{\partial \delta \bfq}{\partial t}
          +\nabla \cdot \delta \pi &=&0, \\
n_0\frac{\partial \delta s}{\partial t}+\frac{1}{T_0}\nabla \cdot \delta \bfq
 &=&0, \\
\delta \Pi +\zeta[\nabla \cdot \delta \bfv 
           +\beta_0\frac{\partial \delta \Pi}{\partial t}
           -\alpha_0\nabla \cdot \delta \bfq]&=&0, \\
\delta \bfq-\kappa T_0[-\frac{1}{T_0}\nabla(\delta T)
              -\frac{\partial \delta \bfv }{\partial t}
              -\beta_1\frac{\partial \delta \bfq}{\partial t}
              -\alpha_0 \nabla (\delta \Pi)
              -\alpha_1 \nabla \cdot \delta \pi ]&=&0, \\
\delta \pi^{i j}-\eta[\partial^i \delta v^j +\partial^j \delta v^i
            -\frac{2}{3}g^{i j}\nabla \cdot \delta \bfv 
            -2\beta_2\frac{\partial \delta \pi^{i j}}{\partial t},\nonumber \\
            -\alpha_1(\partial^i \delta q^j +\partial^j \delta q^i 
            -\frac{2}{3}g^{i j} \nabla \cdot \delta \bfq)]&=& 0.
\eeq

Performing Fourier-Laplace transformation,
we find for the longitudinal components,
\begin{equation}
A  
  \begin{pmatrix}
  \delta n(\bfk ,z)                           \\
  \delta v_{\parallel}(\bfk ,z)               \\
  \delta T(\bfk ,z)                           \\
  \delta \Pi(\bfk ,z)                         \\
  \delta q_{\parallel}(\bfk ,z)               \\
  \delta \pi_{\parallel \; \parallel}(\bfk ,z) 
  \end{pmatrix}
=
  \begin{pmatrix}
  \delta n(\bfk ,0)                                                                   \\
  \beta_0\zeta\delta\Pi(\bfk ,0)                                                     \\
  -(c_s^2 \alpha_P/n_0\gamma)\delta n(\bfk ,0)+(n_0\tilde{c}_n/w_0 T_0)\delta T(\bfk ,0)   \\
  \kappa T_0 \delta v_{\parallel}(\bfk ,0)+\beta_1\kappa T_0 \delta q_{\parallel}(\bfk ,0) \\
  \delta v_{\parallel}(\bfk ,0)+(1/w_0)\delta q_{\parallel}(\bfk ,0)                       \\
  2\eta\beta_2\delta\pi_{\parallel \; \parallel}(\bfk ,0)
  \end{pmatrix},
\end{equation}
where $\delta q_{\parallel}=\bfk \cdot \delta\bfq$, 
$\delta \pi_{\parallel \; \parallel}=k^i \cdot k^j \cdot\pi_{i j}$ and 
\begin{equation}
A=
  \begin{pmatrix}
  z  &  i k n_0 & 0 & 0 & 0 & 0  \\
  0  & i k \zeta & 0 & 1+z\beta_0\zeta &-i k \alpha_0 \zeta & 0 \\ 
  -z\frac{c_s^2 \alpha_P}{n_0 \gamma} & 0 & z\frac{n_0 \tilde{c}_n}{w_0 T_0} & 0 & 
\frac{i k}
{w_0 T_0} & 0\\
    0 & z\kappa T_0 & k\kappa & k\alpha_0\kappa T_0 &
1+z\beta_1\kappa T_0 &k\alpha_1 \\
  k\frac{c_s^2}{n_0\gamma} & z & k\frac{c_s^2\alpha_P}{\gamma} & \frac{k}{w_0} & 
\frac{z}
{w_0} & 
\frac{k}{w_0}   \\  
  0 & \frac{4}{3}k\eta & 0 & 0 & -\frac{4}{3}k\alpha_1\eta & 1+2\beta_2\eta 
\end{pmatrix}.
\end{equation}
For simplicity, we have neglected the coupling terms $\alpha_0$ and $\alpha_1$,
which are expected small and 
irrelevant for the following argument.

Applying the similar argument done in the case of Landau equation, we have
\begin{equation}
\delta n(\bfk ,z)=[(A^{-1})_{1 1}
                     -\frac{c_s^2\alpha_P}{n_0 \gamma}(A^{-1})_{1 3}]\delta n(\bfk ,0).
\end{equation} 
In the present case, $\det A$ is given by
\begin{eqnarray}
  \det A=F_1 z^6+F_2 z^5+F_3 z^4+F_4 z^3
  &&+k^2(F_5 z^4+F_6 z^3+F_7 z^2+ F_8 z)\nonumber \\
  &&+k^4(F_9 z^2+F_{10}z+F_{11})],
\end{eqnarray}
where $F_1 \sim F_{11}$ are defined by
\beq
F_1 &= & 2\beta_0\beta_2\zeta\eta\kappa T_0(\beta_1 w_0-1), \\
F_2 &= & 2_0\beta_0\beta_2\zeta\eta+\kappa T_0(\beta_1 w_0-1)
(\beta_0\zeta+2\beta_2\eta), \\
F_3 &= & \kappa T_0 (\beta_1 w_0-1)+w_0(\beta_0\zeta+2\beta_2\eta), \\
F_4 &= & w_0,\\ 
F_5 &= & 
\beta_1\zeta\eta\kappa T_0(\frac{4}{3}\beta_0+2\beta_2)
 +2\beta_0\beta_2\zeta\eta w_0(\gamma\chi-2\chi c_s^2\alpha_P T_0
+c_s^2\beta_1\kappa  T_0),\\
  F_6 &= &\beta_1\kappa T_0 w_0\nu_l
     +w_0(\beta_0 \nonumber \\
    & &      +2\beta_2\eta)(\gamma\chi-2\chi c_s^2\alpha_P T_0
+c_s^2\beta_1\kappa T_0) \nonumber \\
     & & + \zeta\eta(\frac{4}{3}\beta_0+2\beta_2+2 w_0 c_s^2 \beta_0\beta_2),\\
F_7 & = & 
w_0[\gamma\chi -2\chi c_s^2 \alpha_P T_0+c_s^2 \beta_1 \kappa T_0+\nu_l
        +c_s^2(\beta_0+2\beta_2\eta) ], \\
F_8&=&w_0 c_s^2, \\
F_9 &= & \gamma\chi(\beta_0\zeta+2\beta_2\eta)
         +2\beta_0\beta_2\zeta\eta w_0 c_s^2\chi, \\
F_{10} & = & w_0[\gamma\chi\nu_l+c_s^2\chi(\beta_0\zeta+2\beta_2\eta)], \\
F_{11} &= & w_0 c_s^2 \chi,
\eeq
which are all independently of $k$ and $z$,
As before, $\det A$ can be approximately factorized for small $k$,
\begin{eqnarray}
  \det A \sim \frac{n_0 \tilde{c}_n}{w_0^2 T_0}&&
              (\beta_0\zeta z+1+O(k^2))
              [(\beta_1 w_0-1)\kappa T_0+w_0+O(k^2)]\nonumber \\
              &&\times (2\beta_2\eta z+1+O(k^2)) 
              (z+\chi k^2) \nonumber \\
              &&\times
              (z+\Gamma_{\rm B} k^2+i c_s k)
              (z+\Gamma_{\rm B} k^2-i c_s k),
\end{eqnarray}
with 
\begin{equation}
\Gamma_{\rm B}=\frac{1}{2}[(\gamma-1)\chi+\nu_l +c_s^2 T_0(\kappa /w_0-2\chi\alpha_P )]
\equiv \Gamma^{\rm IS}_{\rm B},
\end{equation}
which will be found to be the width of the Brillouin peaks and is equal to 
that for the Eckart equation, $\Gamma^{\rm Ec}_{\rm B}$.
One can see that the relativistic effect enter two ways; one is due to 
thermal conductivity which tends to enhance the width while
the thermal expansivity would reduce the width. The fact that 
the  relativistic effects 
might possibly enhance the Brillouin width is in contrast
to the case of TKO equation where the relativistic effects only tends
to reduce the width of the sound modes.
As we will see in the text,
the net relativistic contribution to  $\Gamma^{\rm IS}_{\rm B}$ is found to
be negative.

\end{document}